# LINEAR DEPENDENCE OF $\beta^-$-DECAY MAXIMUM ENERGY ON THE MASS NUMBER $A$ ALONG ISOTOPIC CHAINS FOR $Z$<47

Tolga Yarman,[a,*] Nimet Zaim,[b] Alexander Kholmetskii,[c] Ozan Yarman,[d] Faruk Yarman[e]

[a] *Istanbul Okan University, Istanbul, Türkiye (*Corresponding Author: tyarman@gmail.com)*
[b] *Trakya University, Edirne, Türkiye (Retired on 22.04.2024)*
[c] *Belarus State University, Minsk, Belarus*
[d] *Istanbul University, Istanbul, Türkiye*
[e] *Savronik, Çankaya, Ankara, Türkiye*

**Abstract** We investigate the systematics of the maximum $\beta^-$-decay energy $E$ as a function of the mass number $A$ along isotopic chains with a fixed proton number across $Z$<47. By making use of the available curated nuclear data, we find that, for each fixed $Z$, the decay energy can be described to excellent accuracy by a linear dependence on $A$, provided that even-$A$ and odd-$A$ isotopes are treated separately. This yields two straight-line trends for each element, which are characterized by the slope and intercept parameters that can be systematically tabulated across the studied range. The corresponding fits are remarkably accurate, where the coefficients of determination are typically almost unity. Such an element-by-element empirical regularity does not appear to have been previously tabulated in a compact systematic form in the nuclear physics literature. We hence provide a simple and compact parameterization of $\beta^-$-decay energetics along isotopic chains with respect to our stated scope, whereby the approach at hand may prove useful for the analysis of decay-energy evolution, behavioral classification, and preliminary estimates of $\beta^-$-decay properties. The broader theoretical motivation that initially led us to search for such a regularity is discussed only after the confirmation of our results through experimental data is established.

## 1. INTRODUCTION

The maximum $\beta^-$-decay energy $E$ is a fundamental observable in nuclear physics. It is directly related to mechanisms of atomic disintegration, so much so that it influences decay rates and plays an important role in the evaluation of nuclear structure and even astrophysical modeling. The systematics of $\beta$-decay energetics through either a *positron* or an *electron* emission across isotopic chains therefore constitute an important subject of extensive tabulations and theoretical treatments, including semi-empirical mass formulae,[1,2,3,4,5]

[1] C.F.v. Weizsäcker, 1935, *Z. Physik*, **96**.
[2] H. A. Bethe and R. F. Bacher, 1936, *Rev. Mod. Phys.*, **8** (82).
[3] W. D. Myers and W. J. Swiatecki, 1966, *Nucl. Phys.*, **81** (1).
[4] P. Möller *et al.*, 1995, *At. Data Nucl. Data Tables*, **59** (2).
[5] G. Audi *et al.*, 2017, *Chin. Phys. C*, **41** (3), 030001.

shell-model approaches,[6,7,8,9,10] gross-theory parameterizations,[11,12,13,14] and modern microscopic methods such as QRPA.[15,16,17,18,19]

Despite this extensive literature, a particularly simple empirical regularity that tackles all the radioisotopes in elemental groups appears to have been unnoticed. In the present work, we show that, for each fixed proton number within the range $Z$<47, the maximum $\beta^-$-decay energy $E$ depends linearly on the mass number $A$ when even-$A$ and odd-$A$ isotopes are treated separately. Thus, each isotopic chain is described by two straight lines, one for even-$A$ nuclei and one for odd-$A$ nuclei, with a remarkable match every time.

This result provides a compact and transparent parameterization of $\beta^-$-decay energetics along isotopic chains. The resulting slope and intercept parameters can be tabulated element by element, and may serve as a practical tool for characterizing decay-energy evolution as well as for making preliminary estimates about varying disintegration behavior.

The search for the regularity we shall disclose with this contribution was initially motivated by the Universal Matter Architecture (UMA) scaffolding developed in our earlier publications. [20,21,22,23,24,25,26,27,28] Under this framework, binding-energy variations along an isotopic chain suggested that the related $\beta^-$-decay energies should exhibit a basic linearity with increasing $A$ at fixed $Z$.[29,30] Since we choose to present this article primarily as a study for showing how the available empirical findings match the predictions of a simple formalism, we do not rely on the UMA in what follows; we simply would like to emphasize that it provided the original stimulus for examining the curated data in this way.

We organize our presentation in the following manner: Section 2 lays out the formalistic core of our study; thus, for each element with $Z$<47, we demonstrate that the dependence of a set of corresponding $\beta^-$-decay energies $E$ on the mass number $A$ is outstandingly linear, along

---

[6] M. G. Mayer, 1949, *Phys. Rev.*, **75**.

[7] O. Haxel, J. H. D. Jensen and H. E. Suess, 1949, *Phys. Rev.*, **75**.

[8] B.A. Brown and B.H. Wildenthal, 1988, *Annu. Rev. Nucl. Part. Sci.*, **38**.

[9] E. Caurier *et al.*, 2005, *Rev. Mod. Phys.*, **77**.

[10] B. A. Brown, 2022, *Physics*, **4** (2).

[11] K. Takahashi and M. Yamada, 1969, *Prog. Theor. Phys.*, **41** (6).

[12] T. Tachibana, M. Yamada and Y. Yoshida, 1990, *Prog. Theor. Phys.*, **84** (4).

[13] H. Koura and S. Chiba, 2017, *Phys. Rev. C*, **95**, 064304.

[14] M. Gorchtein and C.-Y. Seng, 2023, *Universe*, **9** (9).

[15] J. Suhonen and O. Civitarese, 1994, *Phys. Rev. C*, **49** (6).

[16] T. Marketin, D. Vretenar and P. Ring, 2007, *Phys. Rev. C*, **75** (2), 024304.

[17] I. Daoutidis and P. Ring, 2011, *Phys. Rev. C.*, **83**, 044303.

[18] R. Oktova, 2020, *J. Phys.: Conf. Ser.*, **1428**, 012013.

[19] F. Minato, Z. Niu and H. Liang, 2022, *Phys. Rev. C*, **106** (2), 024306.

[20] T. Yarman, 2004, *Opt. Spectrosc. Mol. Spectrosc.*, **97** (683).

[21] T. Yarman, 2004, *Opt. Spectrosc. Mol. Spectrosc.*, **97** (691).

[22] T. Yarman, 2013, *Phys. Essays*, **26** (473).

[23] T. Yarman *et al.*, 2011, *Sci. Res. Essays*, **6**, 4478.

[24] T. Yarman *et al.*, 2021, *Mol. Phys.*, **119**, 1871087.

[25] T. Yarman *et al.*, 2016, *Eur. Phys. J. A*, **52** (140).

[26] T. Yarman *et al.*, 2017, *Eur. Phys. J. A*, **53** (1).

[27] T. Yarman *et al.*, 2017, *Int. J. Mod. Phys. E*, **26**, 1750068.

[28] C.B. Marchal *et al.*, *Ann. Phys.*, **454** (2023) 169346.

[29] T. Yarman *et al.*, 2024, *Int. J. Mod. Phys. E*, **33** (8), 2450032.

[30] T. Yarman *et al.*, 2025, *Int. J. Mod. Phys. E*, **34** (6), 2550023.

with separate straight-line trends for *even-A* and *odd-A* isotopes. In Section 3, we explain the corresponding slope and intercept parameters for these lines. Finally, we delve into our conclusions in Section 4.

## 2. THE LINEAR DEPENDENCE OF THE MAXIMUM DECAY ENERGY *E* ON THE MASS NUMBER *A* FOR $\beta^-$-DECAYING RADIOISOTOPES OF THE SAME ELEMENT

For each element in the range $Z<47$, we extracted the evaluated $\beta^-$-decay maximum energies $E$ from the National Nuclear Data Center (NNDC)[31] and Brookhaven databases.[32] Whenever the need arose, conversion between energy (keV) and time units (s) obeyed standard $\beta$-decay treatments.[33] Different data are thus altogether gathered in Table 1.

When plotted as a function of the mass number $A$ along an isotopic chain of any given fixed $Z$, the decay energies reveal an unexpectedly simple pattern; *e.g.*, each chain is accurately described by two linear trends: *one for even-A isotopes* and *one for odd-A isotopes.*

Although the present model is geared to work with decay energetics at the origin instead of at the measurement endpoint, still, we can base our formalism on the curated nuclear data. This is because, even when the endpoint $E$ values should be increased as much as the electron rest energy 0.511 MeV when the *E's* are considered at the moment the electron departs from the nucleus, all the same, the conjectural value of $E$'s at the time of emission is negligible as compared to the measured $E$ values in the laboratory which, in general, exceed 10 MeV.

Our drawings are shown in Figs. 1–4. In all the cases examined, an exceptional linear fit, when applied separately to even and odd mass numbers, provides an excellent description of the isotopic evolution of the decay energy.

The fitted parameters $p$ (intercept) and $q$ (slope) for all elements with $Z<47$ are compiled in Table 2. The quality of such a regularity is furthermore quantified through the coefficient of determination $R^2$ in the said table.[34] Thus, across all isotopic chains, we have the following:

- $R^2$ values are typically very close to unity, if not unity;
- they only rarely fall slightly below 0.95;
- they all reflect the high experimental precision of the evaluated decay-energy data (usually at the 0.1% level or better).

The observed linearity holds across a wide range of isotopic chains, and is robust with respect to the selection of the data sets of concern.

---

[31] Brookhaven National Laboratory, National Nuclear Data Center, NuDat 3.0 Nuclear Structure and Decay Data Navigator (http://www.nndc.bnl.gov/chart).

[32] J. K. Tuli *(date is current),* Nuclear Wallet Cards. National Nuclear Data Center, *Brookhaven National Laboratory, New York.*

[33] D.H. Wilkinson, 1989, *Nucl. Instrum. Methods Phys. Res. A*, **275** (103).

[34] The *coefficient of determination* $R^2$ is, as known, an *overall measurement of the departure of the straight line we draw to the data at hand.* The smaller the summation of the squared errors delineated by the straight line points we draw as referred to the data points, the closer $R^2$ is to unity.

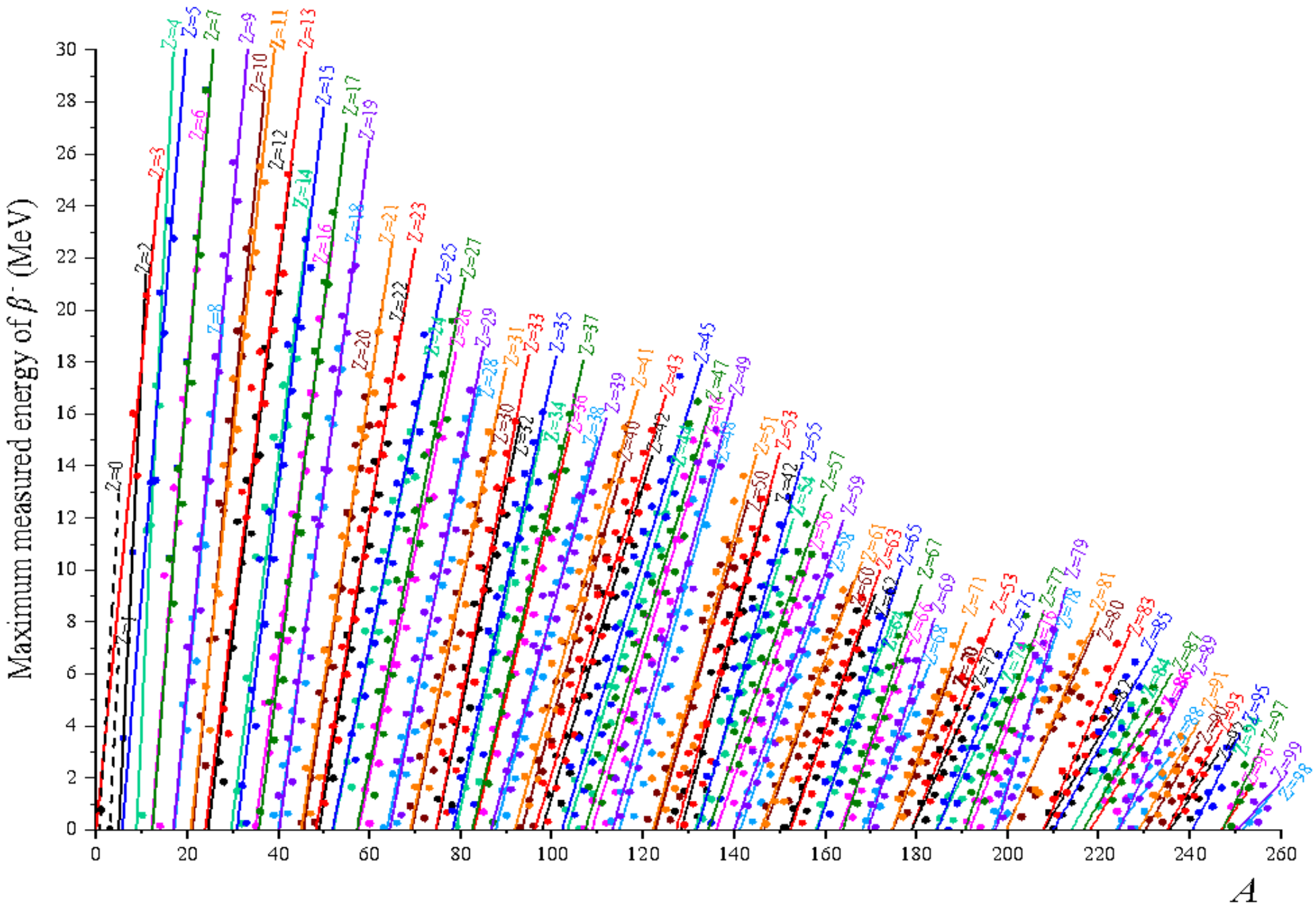

**Figure 1** The fascinating linear dependence, for radioisotopes of the same element per a given *Z*, of the $\beta^-$-decay maximum energy *E* on the mass number *A* across all available *Z's* varying between 0 and 99. We include the range *Z*>47 here for completeness, although it is outside the scope of the current contribution. To avoid overcrowding this panorama, we did not mark the nuclei in the display with their corresponding symbols *(for which, see further below for details)*. The two dashed black lines on the far left indicate *Z*=0 and *Z*=1, where we have only one datapoint for each, and where we have deliberately plotted their lines more or less parallel to the neighboring line at *Z*=2 for visual clarity.

Next, we focus on nuclei whose *Z's* lie in-between *Z*=0 and *Z*=17 (Fig. 2), where we immediately note that the straight lines passing by *even A's (contiguous lines)* and *odd A's (dashed lines)* fit the data better separately.

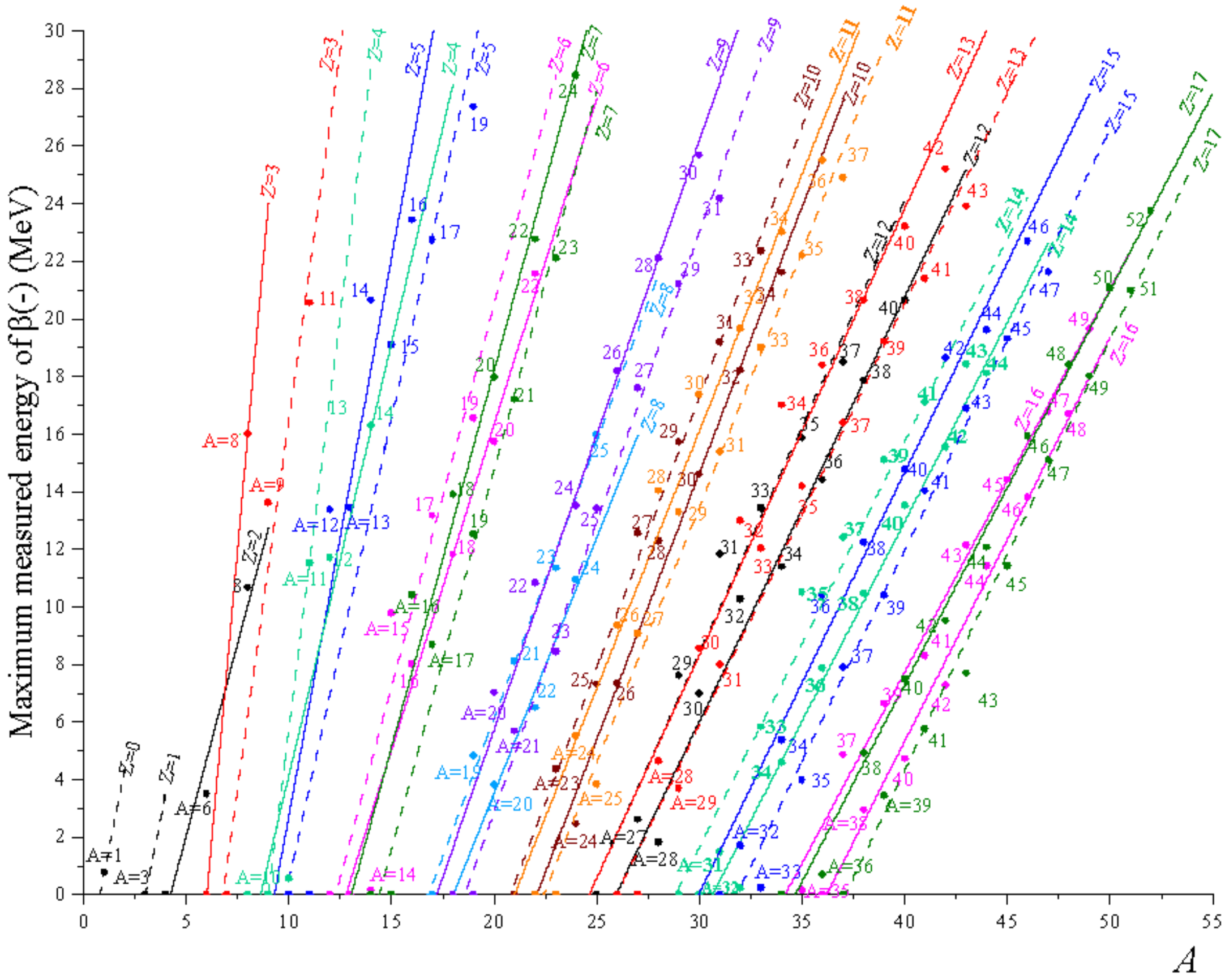

**Figure 2** $\beta^{-}$-decay maximum energy *E* versus the mass number *A* for radioisotopes of the same element per a given *Z* throughout *Z*<18. The contiguous lines trace the *even A* nuclei datapoints, whereas the dashed lines trace the *odd A* nuclei datapoints. Wherever we have just a single nucleus at hand, that is, with respect to *Z*=0 (neutron) and *Z*=1 (Tritium), we have deliberately drawn a line intersecting the single datapoint in question practically parallel to the neighboring straight line at *Z*=3.

This behavior admits a natural explanation:

The vertical offset between the *even-A* and *odd-A* trends reflects the well-known *pairing contribution* to nuclear binding. *Even-A* nuclei being more stable, their decay energies come to be noticeably smaller than those of odd-*A* nuclei within the body of the same element.

The determination of the theoretical confirmation of the *energy gap* on the same vertical between *even-A's and odd-A's* for the same *Z* should, no doubt, be based on the difference of the binding energies per nucleon coming into play in both cases.

In what follows, we plot the maximum decay energy *E* versus *A* for *Z's* lying in-between *Z*=18 and *Z*=33 (Fig. 3).

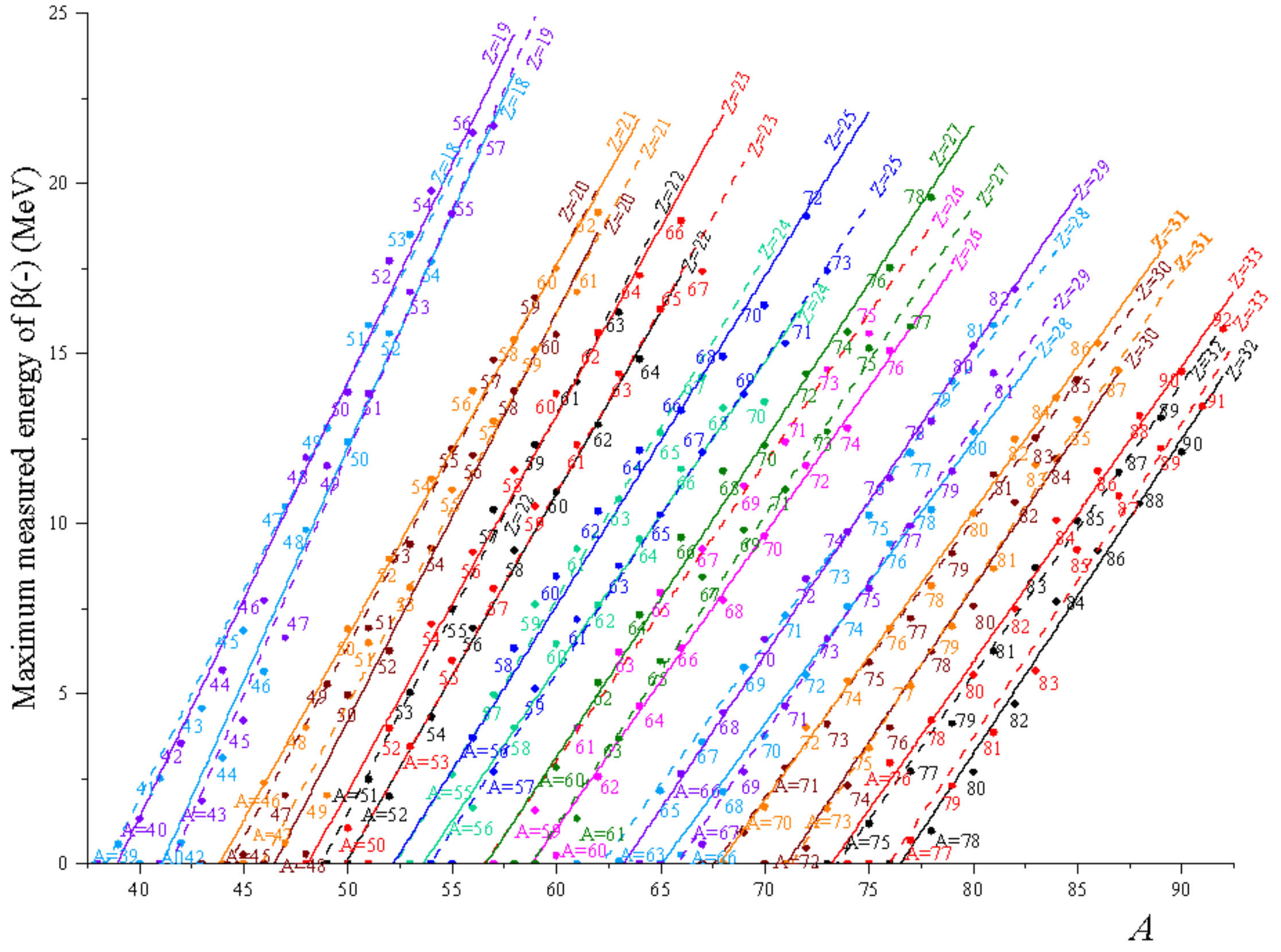

**Figure 3** The maximum $\beta^{-}$-decay energy $E$ versus the mass number $A$ for radioisotopes of the same element in the case of $17<Z<34$. The contiguous straight lines trace the *even A* nuclei datapoints, whereas the dashed lines trace the *odd A* nuclei datapoints.

Proceeding forward, we plot the maximum decay energy $E$ versus $A$ for $Z$*'s* lying in-between $Z$=35 and $Z$=46 (Fig. 4); and this is where we stop—*viz.*, at $Z$=46—for, this presentation is already as voluminous as would be necessary at this point.

Besides, for *higher Z's,* the straight lines of concern become splintered, making it a requisite to draw quadruple lines per any radioisotope ensemble of the same element belonging to a given $Z$ beyond $Z$=46.

Such an endeavor deserves further attention though, which we propose to tackle in a concomitant article.

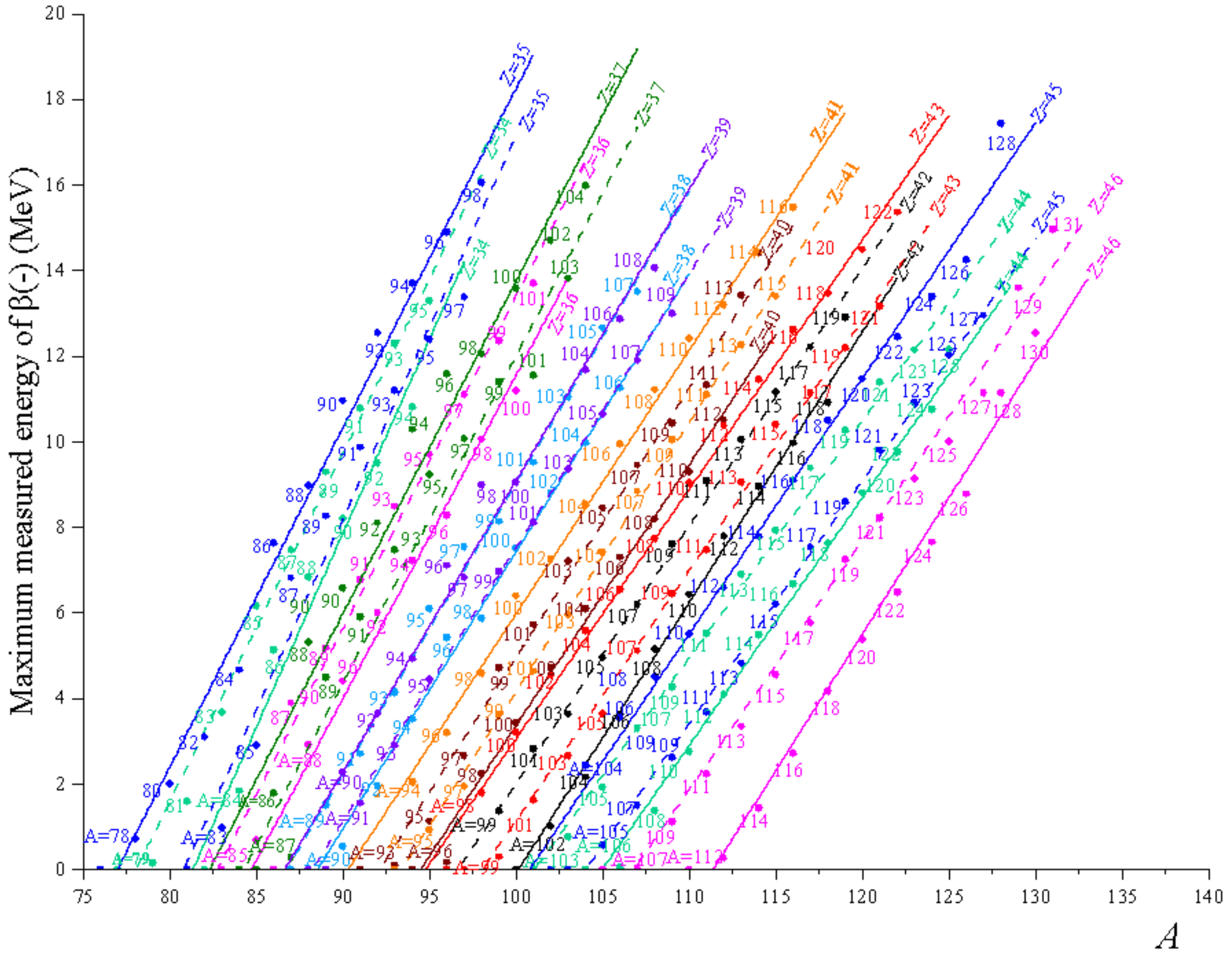

**Figure 4** The maximum $\beta^{-}$-decay energy $E$ versus the mass number $A$ for radioisotopes of the same element in the case of $34<Z<47$. The contiguous straight lines trace the *even A* nuclei datapoints, whereas the dashed lines trace the *odd A* nuclei datapoints.

Provided at this juncture, right below, is Table 1, listing all of the experimentally gathered half-life and decay-energy data pertaining to the radioisotopes encompassed by this study.

**Table 1** Half-lives and decay energies of all radioisotopes considered herein, where those with *unknown* parameters were not plotted in our Figures.

| *Nucleus* | *A* | *Z* | *N* | *T (Half-life) (s)* | *E (Decay Energy) (MeV)* |
|---|---|---|---|---|---|
| n | 1 | 0 | 1 | $6.11 \times 10^{2}$ | 0.78 |
| H | 3 | 1 | 2 | $3.89 \times 10^{8}$ | $1.86 \times 10^{-2}$ |
| He | 6 | 2 | 4 | 0.81 | 3.51 |
| He | 8 | 2 | 6 | 0.12 | 10.66 |
| Li | 8 | 3 | 5 | 0.84 | 16.0 |
| Li | 9 | 3 | 6 | 0.18 | 13.61 |
| Li | 11 | 3 | 8 | $8.80 \times 10^{-3}$ | 20.55 |
| Be | 10 | 4 | 6 | $4.77 \times 10^{13}$ | 0.56 |
| Be | 11 | 4 | 7 | 13.80 | 11.51 |
| Be | 12 | 4 | 8 | $2.15 \times 10^{-2}$ | 11.71 |
| Be | 13 | 4 | 9 | 0.43 | 17.04 |
| Be | 14 | 4 | 10 | $4.35 \times 10^{-3}$ | 16.29 |
| B | 12 | 5 | 7 | $2.0 \times 10^{-3}$ | 13.37 |
| B | 13 | 5 | 8 | $1.70 \times 10^{-2}$ | 13.44 |
| B | 14 | 5 | 9 | $1.25 \times 10^{-2}$ | 20.64 |
| B | 15 | 5 | 10 | $9.90 \times 10^{-3}$ | 19.09 |
| B | 16 | 5 | 11 | <100 keV ($1.82 \times 10^{-17}$s) | 23.42 |
| B | 17 | 5 | 12 | $5.08 \times 10^{-3}$ | 22.74 |
| C | 14 | 6 | 8 | $1.80 \times 10^{11}$ | 0.16 |
| C | 15 | 6 | 9 | 2.45 | 9.77 |
| C | 16 | 6 | 10 | 0.75 | 8.01 |
| C | 17 | 6 | 11 | 0.19 | 13.16 |
| C | 18 | 6 | 12 | $9.20 \times 10^{-2}$ | 11.81 |
| C | 19 | 6 | 13 | $4.62 \times 10^{-2}$ | 16.56 |
| C | 20 | 6 | 14 | $1.62 \times 10^{-2}$ | 15.74 |
| C | 22 | 6 | 16 | $6.10 \times 10^{-3}$ | 21.55 |
| N | 16 | 7 | 9 | 7.13 | 10.42 |
| N | 17 | 7 | 10 | 4.17 | 8.68 |
| N | 18 | 7 | 11 | 0.62 | 13.90 |
| N | 19 | 7 | 12 | 0.34 | 12.52 |
| N | 20 | 7 | 13 | 0.14 | 17.97 |
| N | 21 | 7 | 14 | $8.4 \times 10^{-2}$ | 17.20 |
| N | 22 | 7 | 15 | $1.95 \times 10^{-2}$ | 22.76 |
| N | 23 | 7 | 16 | $1.43 \times 10^{-2}$ | 22.10 |
| N | 24 | 7 | 17 | $5.2 \times 10^{-8}$ | 28.44 |
| O | 19 | 8 | 11 | 26.9 | 4.82 |
| O | 20 | 8 | 12 | 13.50 | 3.81 |
| O | 21 | 8 | 13 | 3.42 | 8.11 |
| O | 22 | 8 | 14 | 2.25 | 6.49 |
| O | 23 | 8 | 15 | $9.70 \times 10^{-2}$ | 11.34 |
| O | 24 | 8 | 16 | $7.20 \times 10^{-2}$ | 10.94 |
| O | 25 | 8 | 17 | 88.0 keV ($6.41 \times 10^{-18}$ s) | 15.98 |
| F | 20 | 9 | 11 | 11.10 | 7.03 |
| F | 21 | 9 | 12 | 4.16 | 5.68 |
| F | 22 | 9 | 13 | 4.23 | 10.82 |
| F | 23 | 9 | 14 | 2.23 | 8.44 |
| F | 24 | 9 | 15 | 0.38 | 13.51 |
| F | 25 | 9 | 16 | $8.0 \times 10^{-2}$ | 13.42 |
| F | 26 | 9 | 17 | $8.20 \times 10^{-3}$ | 18.19 |
| F | 27 | 9 | 18 | $5.0 \times 10^{-3}$ | 17.59 |
| F | 28 | 9 | 19 | 180 keV ($13.10 \times 10^{-18}$ s) | 22.10 |
| F | 29 | 9 | 20 | $2.67 \times 10^{-3}$ | 21.20 |
| F | 30 | 9 | 21 | $<2.50 \times 10^{-7}$ | 25.68 |
| F | 31 | 9 | 22 | *Unknown* | 25.66 |
| Ne | 23 | 10 | 13 | 37.20 | 4.38 |
| Ne | 24 | 10 | 14 | $2.03 \times 10^{2}$ | 2.47 |
| Ne | 25 | 10 | 15 | 0.60 | 7.30 |
| Ne | 26 | 10 | 16 | 0.20 | 7.34 |
| Ne | 27 | 10 | 17 | $3.15 \times 10^{-2}$ | 12.59 |
| Ne | 28 | 10 | 18 | $2.0 \times 10^{-2}$ | 12.28 |
| Ne | 29 | 10 | 19 | $1.50 \times 10^{-2}$ | 15.72 |
| Ne | 30 | 10 | 20 | $7.39 \times 10^{-3}$ | 14.60 |
| Ne | 31 | 10 | 21 | $3.40 \times 10^{-3}$ | 19.18 |
| Ne | 32 | 10 | 22 | $3.50 \times 10^{-3}$ | 18.20 |
| Ne | 33 | 10 | 23 | $<1.80 \times 10^{-7}$ | 22.35 |
| Ne | 34 | 10 | 24 | $<6.0 \times 10^{-8}$ | 21.60 |
| Na | 24 | 11 | 13 | $5.40 \times 10^{4}$ | 5.52 |
| Na | 25 | 11 | 14 | 59.10 | 3.84 |
| Na | 26 | 11 | 15 | 1.07 | 9.35 |
| Na | 27 | 11 | 16 | 0.30 | 9.07 |
| Na | 28 | 11 | 17 | $3.05 \times 10^{-2}$ | 14.03 |
| Na | 29 | 11 | 18 | $4.41 \times 10^{-2}$ | 13.28 |
| Na | 30 | 11 | 19 | $4.80 \times 10^{-2}$ | 17.36 |
| Na | 31 | 11 | 20 | $1.70 \times 10^{-2}$ | 15.38 |
| Na | 32 | 11 | 21 | $1.32 \times 10^{-2}$ | 19.64 |
| Na | 33 | 11 | 22 | $8.0 \times 10^{-3}$ | 19.0 |
| Na | 34 | 11 | 23 | $5.50 \times 10^{-3}$ | 23.0 |
| Na | 35 | 11 | 24 | $1.80 \times 10^{-3}$ | 22.20 |
| Na | 36 | 11 | 25 | $<1.80 \times 10^{-7}$ | 25.50 |
| Na | 37 | 11 | 26 | $>6.0 \times 10^{-8}$ | 24.90 |
| Mg | 27 | 12 | 15 | $5.68 \times 10^{2}$ | 2.61 |
| Mg | 28 | 12 | 16 | $7.53 \times 10^{4}$ | 1.83 |
| Mg | 29 | 12 | 17 | 1.30 | 7.61 |
| Mg | 30 | 12 | 18 | 0.32 | 6.99 |
| Mg | 31 | 12 | 19 | 0.24 | 11.83 |
| Mg | 32 | 12 | 20 | $8.60 \times 10^{-2}$ | 10.27 |
| Mg | 33 | 12 | 21 | $8.94 \times 10^{-2}$ | 13.43 |
| Mg | 34 | 12 | 22 | $2.0 \times 10^{-2}$ | 11.39 |
| Mg | 35 | 12 | 23 | $1.13 \times 10^{-2}$ | 15.86 |
| Mg | 36 | 12 | 24 | $5.80 \times 10^{-3}$ | 14.40 |
| Mg | 37 | 12 | 25 | $8.0 \times 10^{-3}$ | 18.50 |
| Mg | 38 | 12 | 26 | $3.10 \times 10^{-3}$ | 17.86 |
| Mg | 40 | 12 | 28 | $>1.70 \times 10^{-7}$ | 20.63 |
| Al | 28 | 13 | 15 | $1.35 \times 10^{2}$ | 4.64 |
| Al | 29 | 13 | 16 | $3.94 \times 10^{2}$ | 3.69 |
| Al | 30 | 13 | 17 | 3.62 | 8.56 |
| Al | 31 | 13 | 18 | 0.64 | 7.99 |
| Al | 32 | 13 | 19 | $3.19 \times 10^{-2}$ | 12.98 |
| Al | 33 | 13 | 20 | $4.17 \times 10^{-2}$ | 12.02 |
| Al | 34 | 13 | 21 | $5.63 \times 10^{-2}$ | 16.99 |
| Al | 35 | 13 | 22 | $3.76 \times 10^{-2}$ | 14.17 |
| Al | 36 | 13 | 23 | $9.40 \times 10^{-2}$ | 18.39 |
| Al | 37 | 13 | 24 | $1.07 \times 10^{-2}$ | 16.38 |
| Al | 38 | 13 | 25 | $7.60 \times 10^{-3}$ | 20.64 |
| Al | 39 | 13 | 26 | $7.60 \times 10^{-3}$ | 19.2 |
| Al | 40 | 13 | 27 | $5.7 \times 10^{-3}$ ($>2.60 \times 10^{-7}$) | 23.2 |
| Al | 41 | 13 | 28 | $3.5 \times 10^{-3}$ | 21.77 |
| Al | 42 | 13 | 29 | $>1.70 \times 10^{-7}$ | 24.28 |
| Al | 43 | 13 | 30 | $>1.70 \times 10^{-7}$ | 24.84 |
| Si | 31 | 14 | 17 | $9.44 \times 10^{3}$ | 1.4910 |
| Si | 32 | 14 | 18 | $4.83 \times 10^{9}$ | 0.23 |
| Si | 33 | 14 | 19 | 6.11 | 5.82 |
| Si | 34 | 14 | 20 | 2.77 | 4.59 |
| Si | 35 | 14 | 21 | 0.78 | 10.50 |
| Si | 36 | 14 | 22 | 0.45 | 7.86 |
| Si | 37 | 14 | 23 | $9.0 \times 10^{-2}$ | 12.40 |
| Si | 38 | 14 | 24 | $9.50 \times 10^{-2}$ | 10.45 |
| Si | 39 | 14 | 25 | $4.75 \times 10^{-2}$ | 15.09 |
| Si | 40 | 14 | 26 | $3.30 \times 10^{-2}$ | 13.50 |
| Si | 41 | 14 | 27 | $2.0 \times 10^{-2}$ | 17.10 |
| Si | 42 | 14 | 28 | $1.25 \times 10^{-2}$ | 15.55 |
| Si | 43 | 14 | 29 | $1.30 \times 10^{-2}$ ($>6.0 \times 10^{-8}$) | 18.42 |
| Si | 44 | 14 | 30 | *Unknown* | 18.10 |
| P | 32 | 15 | 17 | $1.23 \times 10^{6}$ | 1.71 |
| P | 33 | 15 | 18 | $2.19 \times 10^{6}$ | 0.25 |
| P | 34 | 15 | 19 | 0.12 | 5.38 |
| P | 35 | 15 | 20 | 47.30 | 3.99 |
| P | 36 | 15 | 21 | 5.60 | 10.41 |
| P | 37 | 15 | 22 | 2.31 | 7.90 |
| P | 38 | 15 | 23 | 0.64 | 12.24 |
| P | 39 | 15 | 24 | 0.28 | 10.39 |
| P | 40 | 15 | 25 | 0.15 | 14.76 |
| P | 41 | 15 | 26 | 0.10 | 14.03 |
| P | 42 | 15 | 27 | $4.85 \times 10^{-2}$ | 18.65 |
| P | 43 | 15 | 28 | $3.65 \times 10^{-2}$ | 16.88 |
| P | 44 | 15 | 29 | $1.85 \times 10^{-2}$ | 19.60 |
| P | 45 | 15 | 30 | $2.40 \times 10^{-2}$ ($>2.0 \times 10^{-7}$) | 19.30 |
| P | 46 | 15 | 31 | *Unknown* | 22.70 |
| P | 47 | 15 | 32 | *Unknown* | 21.61 |
| S | 35 | 16 | 19 | $7.55 \times 10^{6}$ | 0.17 |
| S | 37 | 16 | 21 | $3.03 \times 10^{2}$ | 4.87 |
| S | 38 | 16 | 22 | $1.02 \times 10^{4}$ | 2.94 |
| S | 39 | 16 | 23 | 11.50 | 6.64 |
| S | 40 | 16 | 24 | 8.80 | 4.72 |
| S | 41 | 16 | 25 | 2.60 | 8.30 |
| S | 42 | 16 | 26 | 1.02 | 7.28 |
| S | 43 | 16 | 27 | 0.27 | 12.13 |
| S | 44 | 16 | 28 | 0.10 | 11.41 |
| S | 45 | 16 | 29 | $6.80 \times 10^{-2}$ | 14.40 |
| S | 46 | 16 | 30 | $5.0 \times 10^{-2}$ | 13.80 |
| S | 47 | 16 | 31 | *Unknown* | 16.80 |
| S | 48 | 16 | 32 | *Unknown* | 16.70 |
| S | 49 | 16 | 33 | *Unknown* | 19.65 |
| Cl | 36 | 17 | 19 | $9.49 \times 10^{12}$ | 0.71 |
| Cl | 38 | 17 | 21 | $2.23 \times 10^{3}$ | 4.92 |
| Cl | 39 | 17 | 22 | $3.37 \times 10^{3}$ | 3.44 |
| Cl | 40 | 17 | 23 | 810 | 7.48 |
| Cl | 41 | 17 | 24 | 38.4 | 5.76 |
| Cl | 42 | 17 | 25 | 6.80 | 9.51 |
| Cl | 43 | 17 | 26 | 3.18 | 7.69 |
| Cl | 44 | 17 | 27 | 0.65 | 12.06 |
| Cl | 45 | 17 | 28 | 0.40 | 11.41 |
| Cl | 46 | 17 | 29 | 0.23 | 15.92 |
| Cl | 47 | 17 | 30 | 0.10 | 15.10 |
| Cl | 48 | 17 | 31 | *Unknown* | 18.40 |
| Cl | 49 | 17 | 32 | $>1.7 \times 10^{-7}$ | 18.0 |
| Cl | 50 | 17 | 33 | *Unknown* | 21.07 |
| Cl | 51 | 17 | 34 | *Unknown* | 20.98 |
| Cl | 52 | 17 | 35 | *Unknown* | 23.74 |
| Ar | 39 | 18 | 21 | $8.48 \times 10^{9}$ | 0.57 |
| Ar | 41 | 18 | 23 | $6.58 \times 10^{3}$ | 2.49 |
| Ar | 42 | 18 | 24 | $1.04 \times 10^{9}$ | 0.60 |
| Ar | 43 | 18 | 25 | $3.22 \times 10^{2}$ | 4.57 |
| Ar | 44 | 18 | 26 | $7.12 \times 10^{2}$ | 3.11 |
| Ar | 45 | 18 | 27 | 21.50 | 6.84 |
| Ar | 46 | 18 | 28 | 8.40 | 5.68 |
| Ar | 47 | 18 | 29 | 1.23 | 10.50 |
| Ar | 48 | 18 | 30 | 0.42 | 9.80 |
| Ar | 49 | 18 | 31 | 0.24 | 12.80 |
| Ar | 50 | 18 | 32 | 0.11 | 12.40 |
| Ar | 51 | 18 | 33 | *Unknown* | 15.83 |
| Ar | 52 | 18 | 34 | *Unknown* | 15.58 |
| Ar | 53 | 18 | 35 | $>6.20 \times 10^{-7}$ | 18.50 |
| Ar | 54 | 18 | 36 | *Unknown* | 17.71 |
| K | 40 | 19 | 21 | $3.94 \times 10^{16}$ | 1.31 |
| K | 42 | 19 | 23 | $4.45 \times 10^{4}$ | 3.53 |
| K | 43 | 19 | 24 | $8.03 \times 10^{4}$ | 1.83 |
| K | 44 | 19 | 25 | $1.33 \times 10^{3}$ | 5.69 |
| K | 45 | 19 | 26 | $1.07 \times 10^{3}$ | 4.20 |
| K | 46 | 19 | 27 | $1.05 \times 10^{2}$ | 7.72 |
| K | 47 | 19 | 28 | 17.50 | 6.63 |
| K | 48 | 19 | 29 | 6.83 | 11.94 |
| K | 49 | 19 | 30 | 1.26 | 11.69 |
| K | 50 | 19 | 31 | 0.47 | 13.86 |
| K | 51 | 19 | 32 | 0.37 | 13.82 |
| K | 52 | 19 | 33 | 0.11 | 17.72 |
| K | 53 | 19 | 34 | $3.0 \times 10^{-2}$ | 16.80 |
| K | 54 | 19 | 35 | $1.0 \times 10^{-2}$ | 19.78 |
| K | 55 | 19 | 36 | *Unknown* | 19.10 |
| K | 56 | 19 | 37 | *Unknown* | 21.49 |
| K | 57 | 19 | 38 | *Unknown* | 21.69 |
| Ca | 45 | 20 | 25 | $1.41 \times 10^{7}$ | 0.26 |
| Ca | 47 | 20 | 27 | $3.92 \times 10^{5}$ | 1.99 |
| Ca | 48 | 20 | 28 | $9.15 \times 10^{26}$ | 0.28 |
| Ca | 49 | 20 | 29 | $5.23 \times 10^{2}$ | 5.26 |
| Ca | 50 | 20 | 30 | 13.9 | 4.96 |
| Ca | 51 | 20 | 31 | 10.0 | 6.90 |
| Ca | 52 | 20 | 32 | 4.60 | 5.90 |
| Ca | 53 | 20 | 33 | 0.45 | 9.38 |
| Ca | 54 | 20 | 34 | 0.11 | 9.28 |
| Ca | 55 | 20 | 35 | $2.20 \times 10^{-2}$ | 12.19 |
| Ca | 56 | 20 | 36 | $1.10 \times 10^{-2}$ | 12.0 |
| Ca | 57 | 20 | 37 | $>6.20 \times 10^{-7}$ | 14.80 |
| Ca | 58 | 20 | 38 | *Unknown* | 13.90 |
| Ca | 59 | 20 | 39 | *Unknown* | 16.64 |
| Ca | 60 | 20 | 40 | *Unknown* | 15.55 |
| Sc | 46 | 21 | 25 | $7.24 \times 10^{6}$ | 2.366 |
| Sc | 47 | 21 | 26 | $2.90 \times 10^{5}$ | 0.6 |
| Sc | 48 | 21 | 27 | $1.57 \times 10^{5}$ | 3.99 |
| Sc | 49 | 21 | 28 | $3.43 \times 10^{3}$ | 2.0 |
| Sc | 50 | 21 | 29 | $1.03 \times 10^{2}$ | 6.88 |
| Sc | 51 | 21 | 30 | $1.24 \times 10^{1}$ | 6.50 |
| Sc | 52 | 21 | 31 | 0.20 | 9.30 |
| Sc | 53 | 21 | 32 | 2.40 | 8.72 |
| Sc | 54 | 21 | 33 | 0.53 | 12.0 |
| Sc | 55 | 21 | 34 | $9.60 \times 10^{-2}$ | 10.99 |
| Sc | 56 | 21 | 35 | $3.10 \times 10^{-2}$ | 13.91 |
| Sc | 57 | 21 | 36 | $2.0 \times 10^{-3}$ | 13.20 |
| Sc | 58 | 21 | 37 | $1.20 \times 10^{-2}$ | 16.20 |
| Sc | 59 | 21 | 38 | *Unknown* | 15.21 |
| Sc | 60 | 21 | 39 | *Unknown* | 17.50 |
| Sc | 61 | 21 | 40 | *Unknown* | 16.90 |
| Sc | 62 | 21 | 41 | *Unknown* | 19.50 |
| Ti | 51 | 22 | 29 | $3.45 \times 10^{2}$ | 2.47 |
| Ti | 52 | 22 | 30 | $1.02 \times 10^{2}$ | 1.98 |
| Ti | 53 | 22 | 31 | 32.70 | 5.02 |
| Ti | 54 | 22 | 32 | 2.10 | 4.30 |
| Ti | 55 | 22 | 33 | 1.30 | 7.29 |
| Ti | 56 | 22 | 34 | 0.20 | 6.92 |
| Ti | 57 | 22 | 35 | $9.80 \times 10^{-2}$ | 10.4 |
| Ti | 58 | 22 | 36 | $5.80 \times 10^{-2}$ | 9.21 |
| Ti | 59 | 22 | 37 | $2.85 \times 10^{-2}$ | 12.32 |
| Ti | 60 | 22 | 38 | $2.20 \times 10^{-2}$ | 10.91 |
| Ti | 61 | 22 | 39 | $1.50 \times 10^{-2}$ | 14.16 |
| Ti | 62 | 22 | 40 | *Unknown* | 12.90 |
| Ti | 63 | 22 | 41 | *Unknown* | 16.20 |
| Ti | 64 | 22 | 42 | *Unknown* | 14.84 |
| V | 50 | 23 | 27 | $6.62 \times 10^{24}$ | 1.04 |
| V | 52 | 23 | 29 | $2.25 \times 10^{2}$ | 3.97 |
| V | 53 | 23 | 30 | 92.60 | 3.44 |
| V | 54 | 23 | 31 | 49.80 | 7.04 |
| V | 55 | 23 | 32 | 6.54 | 5.97 |
| V | 56 | 23 | 33 | 0.22 | 9.16 |
| V | 57 | 23 | 34 | 0.35 | 8.09 |
| V | 58 | 23 | 35 | 0.19 | 11.56 |
| V | 59 | 23 | 36 | $9.20 \times 10^{-2}$ | 10.51 |
| V | 60 | 23 | 37 | 0.12 | 13.26 |
| V | 61 | 23 | 38 | $4.80 \times 10^{-2}$ | 11.99 |
| V | 62 | 23 | 39 | $3.35 \times 10^{-2}$ | 15.40 |
| V | 63 | 23 | 40 | $1.96 \times 10^{-2}$ | 14.44 |
| V | 64 | 23 | 41 | $1.50 \times 10^{-2}$ | 17.29 |
| V | 65 | 23 | 42 | *Unknown* | 16.3 |
| V | 66 | 23 | 43 | $>3.60 \times 10^{-7}$ | 18.90 |
| V | 67 | 23 | 44 | $>6.20 \times 10^{-7}$ | 17.41 |
| Cr | 55 | 24 | 31 | $2.10 \times 10^{2}$ | 2.60 |
| Cr | 56 | 24 | 32 | $3.56 \times 10^{2}$ | 1.63 |
| Cr | 57 | 24 | 33 | 21.10 | 4.96 |
| Cr | 58 | 24 | 34 | 7.0 | 3.99 |
| Cr | 59 | 24 | 35 | 1.05 | 7.63 |
| Cr | 60 | 24 | 36 | 0.49 | 6.46 |
| Cr | 61 | 24 | 37 | 0.23 | 9.25 |
| Cr | 62 | 24 | 38 | 0.20 | 7.59 |
| Cr | 63 | 24 | 39 | 0.13 | 10.70 |
| Cr | 64 | 24 | 40 | $4.30 \times 10^{-2}$ | 9.53 |
| Cr | 65 | 24 | 41 | $2.80 \times 10^{-2}$ | 12.66 |
| Cr | 66 | 24 | 42 | $2.30 \times 10^{-2}$ | 11.60 |
| Cr | 67 | 24 | 43 | *Unknown* | 14.30 |
| Cr | 68 | 24 | 44 | $>3.60 \times 10^{-7}$ | 13.40 |
| Cr | 69 | 24 | 45 | $>6.20 \times 10^{-7}$ | 15.29 |
| Cr | 70 | 24 | 46 | $>6.20 \times 10^{-7}$ | 13.58 |
| Mn | 56 | 25 | 31 | $2.228 \times 10^{5}$ | 3.70 |
| Mn | 57 | 25 | 32 | 85.4 | 2.70 |
| Mn | 58 | 25 | 33 | 3.0 | 6.33 |
| Mn | 59 | 25 | 34 | 4.59 | 5.14 |
| Mn | 60 | 25 | 35 | 0.28 | 8.44 |
| Mn | 61 | 25 | 36 | 0.71 | 7.18 |
| Mn | 62 | 25 | 37 | $9.20 \times 10^{-2}$ | 10.35 |
| Mn | 63 | 25 | 38 | 0.28 | 8.75 |
| Mn | 64 | 25 | 39 | $9.0 \times 10^{-2}$ | 12.15 |
| Mn | 65 | 25 | 40 | $9.20 \times 10^{-2}$ | 10.25 |
| Mn | 66 | 25 | 41 | $4.0 \times 10^{-3}$ | 13.32 |
| Mn | 67 | 25 | 42 | $4.70 \times 10^{-2}$ | 12.10 |
| Mn | 68 | 25 | 43 | $2.80 \times 10^{-2}$ | 14.90 |
| Mn | 69 | 25 | 44 | $1.60 \times 10^{-2}$ | 13.80 |
| Mn | 70 | 25 | 45 | $1.99 \times 10^{-2}$ | 16.40 |
| Mn | 71 | 25 | 46 | *Unknown* | 15.30 |
| Mn | 72 | 25 | 47 | $>6.20 \times 10^{-7}$ | 19.04 |
| Mn | 73 | 25 | 48 | *Unknown* | 17.43 |
| Fe | 59 | 26 | 33 | $3.84 \times 10^{6}$ | 1.57 |
| Fe | 60 | 26 | 34 | $8.26 \times 10^{13}$ | 0.24 |
| Fe | 61 | 26 | 35 | $3.59 \times 10^{2}$ | 3.98 |
| Fe | 62 | 26 | 36 | 68.0 | 2.55 |
| Fe | 63 | 26 | 37 | 6.10 | 6.22 |
| Fe | 64 | 26 | 38 | 2.0 | 4.62 |
| Fe | 65 | 26 | 39 | 0.18 | 7.96 |
| Fe | 66 | 26 | 40 | 0.35 | 6.34 |
| Fe | 67 | 26 | 41 | 0.40 | 9.25 |
| Fe | 68 | 26 | 42 | 0.19 | 7.75 |
| Fe | 69 | 26 | 43 | 0.16 | 11.10 |
| Fe | 70 | 26 | 44 | $6.3 \times 10^{-2}$ | 9.63 |
| Fe | 71 | 26 | 45 | $3.5 \times 10^{-2}$ | 12.40 |
| Fe | 72 | 26 | 46 | $>1.50 \times 10^{-7}$ ($1.90 \times 10^{-2}$) | 11.70 |
| Fe | 73 | 26 | 47 | $1.29 \times 10^{-2}$ | 14.52 |
| Fe | 74 | 26 | 48 | $2.0 \times 10^{-3}$ | 12.80 |
| Fe | 75 | 26 | 49 | $>6.20 \times 10^{-7}$ | 15.58 |
| Fe | 76 | 26 | 50 | *Unknown* | 15.07 |
| Co | 60 | 27 | 33 | $1.66 \times 10^{8}$ | 2.82 |
| Co | 61 | 27 | 34 | $5.94 \times 10^{3}$ | 1.32 |
| Co | 62 | 27 | 35 | 92.40 | 5.32 |
| Co | 63 | 27 | 36 | 27.40 | 3.66 |
| Co | 64 | 27 | 37 | 0.30 | 7.31 |
| Co | 65 | 27 | 38 | 1.16 | 5.94 |
| Co | 66 | 27 | 39 | 0.21 | 9.60 |
| Co | 67 | 27 | 40 | 0.43 | 8.42 |
| Co | 68 | 27 | 41 | 0.20 | 11.54 |
| Co | 69 | 27 | 42 | 0.23 | 9.81 |
| Co | 70 | 27 | 43 | 0.11 | 12.29 |
| Co | 71 | 27 | 44 | 0.08 | 11.0 |
| Co | 72 | 27 | 45 | $5.99 \times 10^{-2}$ | 14.40 |
| Co | 73 | 27 | 46 | $4.10 \times 10^{-2}$ | 12.69 |
| Co | 74 | 27 | 47 | $3.14 \times 10^{-2}$ | 15.64 |
| Co | 75 | 27 | 48 | $3.0 \times 10^{-2}$ | 15.15 |
| Co | 76 | 27 | 49 | $2.20 \times 10^{-2}$ | 17.51 |
| Co | 77 | 27 | 50 | $1.30 \times 10^{-2}$ | 15.79 |
| Co | 78 | 27 | 51 | *Unknown* | 19.56 |
| Ni | 63 | 28 | 35 | $3.19 \times 10^{9}$ | $6.69 \times 10^{-2}$ |
| Ni | 65 | 28 | 37 | $9.06 \times 10^{3}$ | 2.14 |
| Ni | 66 | 28 | 38 | $1.97 \times 10^{5}$ | 0.25 |
| Ni | 67 | 28 | 39 | 21.0 | 3.58 |
| Ni | 68 | 28 | 40 | 29.0 | 2.10 |
| Ni | 69 | 28 | 41 | 11.40 | 5.76 |
| Ni | 70 | 28 | 42 | 6.0 | 3.76 |
| Ni | 71 | 28 | 43 | 2.56 | 7.31 |
| Ni | 72 | 28 | 44 | 1.57 | 5.56 |
| Ni | 73 | 28 | 45 | 0.84 | 8.88 |
| Ni | 74 | 28 | 46 | 0.51 | 7.55 |
| Ni | 75 | 28 | 47 | 0.34 | 10.23 |
| Ni | 76 | 28 | 48 | 0.24 | 9.40 |
| Ni | 77 | 28 | 49 | 0.16 | 12.06 |
| Ni | 78 | 28 | 50 | 0.11 | 10.40 |
| Ni | 79 | 28 | 51 | $4.30 \times 10^{-2}$ | 14.19 |
| Ni | 80 | 28 | 52 | $2.40 \times 10^{-2}$ | 12.70 |
| Ni | 81 | 28 | 53 | *Unknown* | 15.82 |
| Cu | 66 | 29 | 37 | $3.07 \times 10^{2}$ | 2.64 |
| Cu | 67 | 29 | 38 | $2.23 \times 10^{5}$ | 0.56 |
| Cu | 68 | 29 | 39 | 30.90 | 4.44 |
| Cu | 69 | 29 | 40 | $1.71 \times 10^{2}$ | 2.68 |
| Cu | 70 | 29 | 41 | 44.50 | 6.59 |
| Cu | 71 | 29 | 42 | 19.40 | 4.62 |
| Cu | 72 | 29 | 43 | 6.63 | 8.36 |
| Cu | 73 | 29 | 44 | 4.20 | 6.61 |
| Cu | 74 | 29 | 45 | 1.63 | 9.75 |
| Cu | 75 | 29 | 46 | 1.22 | 8.09 |
| Cu | 76 | 29 | 47 | 0.64 | 11.33 |
| Cu | 77 | 29 | 48 | 0.47 | 9.93 |
| Cu | 78 | 29 | 49 | 0.34 | 13.0 |
| Cu | 79 | 29 | 50 | 0.24 | 11.53 |
| Cu | 80 | 29 | 51 | 0.11 | 15.22 |
| Cu | 81 | 29 | 52 | $7.30 \times 10^{-2}$ | 14.41 |
| Cu | 82 | 29 | 53 | $3.30 \times 10^{-2}$ | 16.90 |
| Zn | 69 | 30 | 39 | $3.38 \times 10^{3}$ | 0.91 |
| Zn | 71 | 30 | 41 | $1.47 \times 10^{2}$ | 2.81 |
| Zn | 72 | 30 | 42 | $1.67 \times 10^{5}$ | 0.44 |
| Zn | 73 | 30 | 43 | 23.50 | 4.11 |
| Zn | 74 | 30 | 44 | 95.60 | 2.29 |
| Zn | 75 | 30 | 45 | 10.20 | 5.91 |
| Zn | 76 | 30 | 46 | 5.70 | 3.99 |
| Zn | 77 | 30 | 47 | 2.08 | 7.20 |
| Zn | 78 | 30 | 48 | 1.47 | 6.22 |
| Zn | 79 | 30 | 49 | 0.75 | 9.12 |
| Zn | 80 | 30 | 50 | 0.56 | 7.58 |
| Zn | 81 | 30 | 51 | 0.32 | 11.43 |
| Zn | 82 | 30 | 52 | 0.17 | 10.62 |
| Zn | 83 | 30 | 53 | 0.12 | 12.52 |
| Zn | 84 | 30 | 54 | $5.30 \times 10^{-2}$ | 11.90 |
| Zn | 85 | 30 | 55 | $>6.37 \times 10^{-7}$ | 14.22 |
| Ga | 70 | 31 | 39 | $1.27 \times 10^{3}$ | 1.65 |
| Ga | 72 | 31 | 41 | $5.08 \times 10^{4}$ | 4.0 |
| Ga | 73 | 31 | 42 | $1.75 \times 10^{4}$ | 1.6 |
| Ga | 74 | 31 | 43 | $4.87 \times 10^{2}$ | 5.37 |
| Ga | 75 | 31 | 44 | $1.26 \times 10^{2}$ | 3.39 |
| Ga | 76 | 31 | 45 | 32.60 | 6.92 |
| Ga | 77 | 31 | 46 | 13.20 | 5.22 |
| Ga | 78 | 31 | 47 | 5.09 | 8.16 |
| Ga | 79 | 31 | 48 | 2.85 | 6.98 |
| Ga | 80 | 31 | 49 | 1.68 | 10.31 |
| Ga | 81 | 31 | 50 | 1.22 | 8.66 |
| Ga | 82 | 31 | 51 | 0.60 | 12.48 |
| Ga | 83 | 31 | 52 | 0.31 | 11.72 |
| Ga | 84 | 31 | 53 | $8.50 \times 10^{-2}$ | 13.69 |
| Ga | 85 | 31 | 54 | $9.20 \times 10^{-2}$ | 13.06 |
| Ga | 86 | 31 | 55 | $4.30 \times 10^{-2}$ | 15.30 |
| Ga | 87 | 31 | 56 | $2.90 \times 10^{-2}$ ($>6.34 \times 10^{-7}$) | 14.50 |
| Ge | 75 | 32 | 43 | $4.97 \times 10^3$ | 1.18 |
| Ge | 77 | 32 | 45 | $4.04 \times 10^4$ | 2.70 |
| Ge | 78 | 32 | 46 | $5.28 \times 10^3$ | 0.96 |
| Ge | 79 | 32 | 47 | 18.98 | 4.11 |
| Ge | 80 | 32 | 48 | 29.50 | 2.68 |
| Ge | 81 | 32 | 49 | 7.60 | 6.24 |
| Ge | 82 | 32 | 50 | 4.0 | 4.69 |
| Ge | 83 | 32 | 51 | 1.85 | 8.69 |
| Ge | 84 | 32 | 52 | 0.95 | 7.71 |
| Ge | 85 | 32 | 53 | 0.50 | 10.07 |
| Ge | 86 | 32 | 54 | 0.23 | 9.20 |
| Ge | 87 | 32 | 55 | 0.14 | 11.50 |
| Ge | 88 | 32 | 56 | $>3.0 \times 10^{-7}$ ($6.1 \times 10^{-2}$) | 10.58 |
| Ge | 89 | 32 | 57 | *Unknown* | 13.10 |
| Ge | 90 | 32 | 58 | *Unknown* | 12.10 |
| As | 76 | 33 | 43 | $9.46 \times 10^4$ | 2.96 |
| As | 77 | 33 | 44 | $1.40 \times 10^5$ | 0.68 |
| As | 78 | 33 | 45 | $5.44 \times 10^3$ | 4.21 |
| As | 79 | 33 | 46 | $5.41 \times 10^2$ | 2.28 |
| As | 80 | 33 | 47 | 15.20 | 5.55 |
| As | 81 | 33 | 48 | 33.3 | 3.86 |
| As | 82 | 33 | 49 | 19.10 | 7.49 |
| As | 83 | 33 | 50 | 13.40 | 5.67 |
| As | 84 | 33 | 51 | 4.02 | 10.09 |
| As | 85 | 33 | 52 | 2.02 | 9.22 |
| As | 86 | 33 | 53 | 0.95 | 11.54 |
| As | 87 | 33 | 54 | 0.48 | 10.81 |
| As | 88 | 33 | 55 | 0.20 | 13.16 |
| As | 89 | 33 | 56 | *Unknown* | 12.20 |
| As | 90 | 33 | 57 | *Unknown* | 14.47 |
| As | 91 | 33 | 58 | *Unknown* | 13.44 |
| As | 92 | 33 | 59 | *Unknown* | 15.70 |
| Se | 79 | 34 | 45 | $1.03 \times 10^{13}$ | 0.15 |
| Se | 81 | 34 | 47 | $1.11 \times 10^3$ | 1.59 |
| Se | 83 | 34 | 49 | $1.34 \times 10^{3}$ | 3.67 |
| Se | 84 | 34 | 50 | $1.96 \times 10^{2}$ | 1.84 |
| Se | 85 | 34 | 51 | 32.90 | 6.16 |
| Se | 86 | 34 | 52 | 14.30 | 5.13 |
| Se | 87 | 34 | 53 | 5.50 | 7.47 |
| Se | 88 | 34 | 54 | 1.53 | 6.83 |
| Se | 89 | 34 | 55 | 0.43 | 9.28 |
| Se | 90 | 34 | 56 | 0.20 | 8.20 |
| Se | 91 | 34 | 57 | 0.27 | 10.77 |
| Se | 92 | 34 | 58 | *Unknown* | 9.50 |
| Se | 93 | 34 | 59 | *Unknown* | 12.30 |
| Se | 94 | 34 | 60 | $1.50 \times 10^{-7}$ | 10.80 |
| Se | 95 | 34 | 61 | $3.92 \times 10^{-7}$ | 13.30 |
| Br | 78 | 35 | 43 | $3.87 \times 10^{2}$ | 0.73 |
| Br | 80 | 35 | 45 | $1.06 \times 10^{3}$ | 2.0 |
| Br | 82 | 35 | 47 | $1.27 \times 10^{5}$ | 3.09 |
| Br | 83 | 35 | 48 | $8.55 \times 10^{3}$ | 0.98 |
| Br | 84 | 35 | 49 | $1.91 \times 10^{3}$ | 4.66 |
| Br | 85 | 35 | 50 | $1.74 \times 10^{2}$ | 2.91 |
| Br | 86 | 35 | 51 | 55.10 | 7.63 |
| Br | 87 | 35 | 52 | 55.65 | 6.82 |
| Br | 88 | 35 | 53 | 16.34 | 8.98 |
| Br | 89 | 35 | 54 | 4.36 | 8.26 |
| Br | 90 | 35 | 55 | 1.92 | 10.96 |
| Br | 91 | 35 | 56 | 0.54 | 9.87 |
| Br | 92 | 35 | 57 | 0.31 | 12.54 |
| Br | 93 | 35 | 58 | 0.10 | 11.20 |
| Br | 94 | 35 | 59 | 0.07 | 13.70 |
| Br | 95 | 35 | 60 | $1.50 \times 10^{-7}$ | 12.39 |
| Br | 96 | 35 | 61 | $1.50 \times 10^{-7}$ | 14.90 |
| Br | 97 | 35 | 62 | $3.0 \times 10^{-7}$ | 13.37 |
| Br | 98 | 35 | 63 | $6.34 \times 10^{-7}$ | 16.06 |
| Kr | 85 | 36 | 49 | $3.39 \times 10^{8}$ | 0.69 |
| Kr | 87 | 36 | 51 | $4.58 \times 10^{3}$ | 3.89 |
| Kr | 88 | 36 | 52 | $1.02 \times 10^{4}$ | 2.92 |
| Kr | 89 | 36 | 53 | $1.89 \times 10^{2}$ | 5.18 |
| Kr | 90 | 36 | 54 | 32.32 | 4.41 |
| Kr | 91 | 36 | 55 | 8.57 | 6.77 |
| Kr | 92 | 36 | 56 | 1.84 | 6.0 |
| Kr | 93 | 36 | 57 | 1.29 | 8.49 |
| Kr | 94 | 36 | 58 | 0.21 | 7.22 |
| Kr | 95 | 36 | 59 | 0.11 | 9.70 |
| Kr | 96 | 36 | 60 | 0.08 | 8.28 |
| Kr | 97 | 36 | 61 | 0.06 | 11.10 |
| Kr | 98 | 36 | 62 | 0.04 | 10.06 |
| Kr | 99 | 36 | 63 | 0.01 | 12.36 |
| Kr | 100 | 36 | 64 | 0.01 | 11.20 |
| Kr | 101 | 36 | 65 | $>6.35 \times 10^{-7}$ | 13.70 |
| Rb | 86 | 37 | 49 | $1.61 \times 10^{6}$ | 1.78 |
| Rb | 87 | 37 | 50 | $1.57 \times 10^{18}$ | 0.28 |
| Rb | 88 | 37 | 51 | $1.07 \times 10^{3}$ | 5.31 |
| Rb | 89 | 37 | 52 | $9.19 \times 10^{2}$ | 4.50 |
| Rb | 90 | 37 | 53 | $1.58 \times 10^{2}$ | 6.58 |
| Rb | 91 | 37 | 54 | 58.20 | 5.91 |
| Rb | 92 | 37 | 55 | 4.48 | 8.10 |
| Rb | 93 | 37 | 56 | 5.84 | 7.47 |
| Rb | 94 | 37 | 57 | 2.70 | 10.28 |
| Rb | 95 | 37 | 58 | 0.38 | 9.23 |
| Rb | 96 | 37 | 59 | 0.20 | 11.58 |
| Rb | 97 | 37 | 60 | 0.17 | 10.06 |
| Rb | 98 | 37 | 61 | 0.10 | 12.05 |
| Rb | 99 | 37 | 62 | 0.05 | 11.40 |
| Rb | 100 | 37 | 63 | 0.05 | 13.57 |
| Rb | 101 | 37 | 64 | 0.03 | 11.54 |
| Rb | 102 | 37 | 65 | 0.04 | 14.70 |
| Rb | 103 | 37 | 66 | 0.02 | 13.81 |
| Rb | 104 | 37 | 67 | *Unknown* | 15.98 |
| Sr | 89 | 38 | 51 | $4.37 \times 10^{6}$ | 1.50 |
| Sr | 90 | 38 | 52 | $9.11 \times 10^{8}$ | 0.55 |
| Sr | 91 | 38 | 53 | $3.47 \times 10^{4}$ | 2.70 |
| Sr | 92 | 38 | 54 | $9.58 \times 10^{3}$ | 1.95 |
| Sr | 93 | 38 | 55 | $4.46 \times 10^{2}$ | 4.14 |
| Sr | 94 | 38 | 56 | 75.30 | 3.51 |
| Sr | 95 | 38 | 57 | 23.90 | 6.09 |
| Sr | 96 | 38 | 58 | 1.07 | 5.42 |
| Sr | 97 | 38 | 59 | 0.43 | 7.55 |
| Sr | 98 | 38 | 60 | 0.65 | 5.87 |
| Sr | 99 | 38 | 61 | 0.27 | 8.13 |
| Sr | 100 | 38 | 62 | 0.20 | 7.5 |
| Sr | 101 | 38 | 63 | 0.12 | 9.51 |
| Sr | 102 | 38 | 64 | 0.07 | 8.810 |
| Sr | 103 | 38 | 65 | $5.30 \times 10^{-2}$ | 11.04 |
| Sr | 104 | 38 | 66 | $5.30 \times 10^{-2}$ | 9.96 |
| Sr | 105 | 38 | 67 | $3.90 \times 10^{-2}$ | 12.66 |
| Sr | 106 | 38 | 68 | $2.0 \times 10^{-2}$ | 11.26 |
| Sr | 107 | 38 | 69 | $>3.95 \times 10^{-7}$ | 13.50 |
| Y | 90 | 39 | 51 | $2.31 \times 10^{5}$ | 2.28 |
| Y | 91 | 39 | 52 | $5.06 \times 10^{6}$ | 1.54 |
| Y | 92 | 39 | 53 | $1.27 \times 10^{4}$ | 3.64 |
| Y | 93 | 39 | 54 | $3.66 \times 10^{4}$ | 2.90 |
| Y | 94 | 39 | 55 | $1.12 \times 10^{3}$ | 4.92 |
| Y | 95 | 39 | 56 | $6.18 \times 10^{2}$ | 4.45 |
| Y | 96 | 39 | 57 | 5.34 | 7.10 |
| Y | 97 | 39 | 58 | 3.75 | 6.8 |
| Y | 98 | 39 | 59 | 0.55 | 8.992 |
| Y | 99 | 39 | 60 | 1.48 | 6.97 |
| Y | 100 | 39 | 61 | 0.73 | 9.05 |
| Y | 101 | 39 | 62 | 0.45 | 8.10 |
| Y | 102m | 39 | 63 | 0.36 | 10.42 |
| Y | 103 | 39 | 64 | 0.23 | 9.36 |
| Y | 104 | 39 | 65 | 0.20 | 11.67 |
| Y | 105 | 39 | 66 | 0.11 | 10.65 |
| Y | 106 | 39 | 67 | 0.08 | 12.86 |
| Y | 107 | 39 | 68 | 0.03 | 11.90 |
| Y | 108 | 39 | 69 | 0.03 | 14.06 |
| Y | 109 | 39 | 70 | 0.03 | 12.99 |
| Zr | 93 | 40 | 53 | $5.08 \times 10^{13}$ | 0.09 |
| Zr | 95 | 40 | 55 | $5.53 \times 10^{6}$ | 1.12 |
| Zr | 96 | 40 | 56 | $7.41 \times 10^{26}$ | 0.16 |
| Zr | 97 | 40 | 57 | $6.03 \times 10^{4}$ | 2.66 |
| Zr | 98 | 40 | 58 | 30.70 | 2.24 |
| Zr | 99 | 40 | 59 | 2.10 | 4.72 |
| Zr | 100 | 40 | 60 | 7.10 | 3.43 |
| Zr | 101 | 40 | 61 | 2.30 | 5.72 |
| Zr | 102 | 40 | 62 | 2.90 | 4.72 |
| Zr | 103 | 40 | 63 | 1.30 | 7.20 |
| Zr | 104 | 40 | 64 | 1.20 | 6.10 |
| Zr | 105 | 40 | 65 | 0.67 | 8.40 |
| Zr | 106 | 40 | 66 | 0.18 | 7.29 |
| Zr | 107 | 40 | 67 | 0.15 | 9.45 |
| Zr | 108 | 40 | 68 | 0.08 | 8.19 |
| Zr | 109 | 40 | 69 | $5.6 \times 10^{-2}$ | 10.43 |
| Zr | 110 | 40 | 70 | $3.8 \times 10^{-2}$ | 9.30 |
| Zr | 111 | 40 | 71 | $2.4 \times 10^{-2}$ | 11.32 |
| Zr | 112 | 40 | 72 | $3.0 \times 10^{-2}$ | 10.50 |
| Zr | 113 | 40 | 73 | *Unknown* | 13.42 |
| Nb | 94 | 41 | 53 | $6.40 \times 10^{11}$ | 2.05 |
| Nb | 95 | 41 | 54 | $3.02 \times 10^{6}$ | 0.93 |
| Nb | 96 | 41 | 55 | $8.41 \times 10^{4}$ | 3.19 |
| Nb | 97 | 41 | 56 | $4.33 \times 10^{3}$ | 1.94 |
| Nb | 98 | 41 | 57 | 2.86 | 4.59 |
| Nb | 99 | 41 | 58 | 15.0 | 3.64 |
| Nb | 100 | 41 | 59 | 1.40 | 6.40 |
| Nb | 101 | 41 | 60 | 7.10 | 4.63 |
| Nb | 102 | 41 | 61 | 4.30 | 7.26 |
| Nb | 103 | 41 | 62 | 1.50 | 5.94 |
| Nb | 104 | 41 | 63 | 4.90 | 8.53 |
| Nb | 105 | 41 | 64 | 2.91 | 7.42 |
| Nb | 106 | 41 | 65 | 1.02 | 9.94 |
| Nb | 107 | 41 | 66 | 0.30 | 8.84 |
| Nb | 108 | 41 | 67 | 0.20 | 11.22 |
| Nb | 109 | 41 | 68 | 0.11 | 10.05 |
| Nb | 110 | 41 | 69 | $8.20 \times 10^{-2}$ | 12.41 |
| Nb | 111 | 41 | 70 | $5.40 \times 10^{-2}$ | 11.10 |
| Nb | 112 | 41 | 71 | $3.30 \times 10^{-2}$ | 13.20 |
| Nb | 113 | 41 | 72 | $3.20 \times 10^{-2}$ | 12.26 |
| Nb | 114 | 41 | 73 | $1.70 \times 10^{-2}$ | 14.42 |
| Nb | 115 | 41 | 74 | $2.30 \times 10^{-2}$ | 13.40 |
| Nb | 116 | 41 | 75 | *Unknown* | 15.47 |
| Mo | 99 | 42 | 57 | $2.38 \times 10^{5}$ | 1.36 |
| Mo | 101 | 42 | 59 | $8.77 \times 10^{2}$ | 2.82 |
| Mo | 102 | 42 | 60 | $6.78 \times 10^{2}$ | 1.01 |
| Mo | 103 | 42 | 61 | 67.50 | 3.64 |
| Mo | 104 | 42 | 62 | 60.0 | 2.16 |
| Mo | 105 | 42 | 63 | 35.60 | 4.95 |
| Mo | 106 | 42 | 64 | 8.73 | 3.64 |
| Mo | 107 | 42 | 65 | 3.50 | 6.19 |
| Mo | 108 | 42 | 66 | 1.09 | 5.16 |
| Mo | 109 | 42 | 67 | 0.61 | 7.61 |
| Mo | 110 | 42 | 68 | 0.30 | 6.43 |
| Mo | 111 | 42 | 69 | 0.19 | 9.09 |
| Mo | 112 | 42 | 70 | 0.12 | 7.79 |
| Mo | 113 | 42 | 71 | $8.0 \times 10^{-2}$ | 10.04 |
| Mo | 114 | 42 | 72 | $5.80 \times 10^{-2}$ | 8.96 |
| Mo | 115 | 42 | 73 | $4.50 \times 10^{-2}$ | 11.16 |
| Mo | 116 | 42 | 74 | $3.20 \times 10^{-2}$ | 9.96 |
| Mo | 117 | 42 | 75 | $2.20 \times 10^{-2}$ | 12.21 |
| Mo | 118 | 42 | 76 | $1.90 \times 10^{-2}$ | 10.92 |
| Mo | 119 | 42 | 77 | *Unknown* | 12.90 |
| Tc | 98 | 43 | 55 | $1.32 \times 10^{14}$ | 1.80 |

## 3. PARAMETERS OF THE DRAWN STRAIGHT LINES

We now propose to write the maximum energy $E$ of the $\beta^-$-decay under consideration as

$$E = p + qA\ , \quad (1)$$

where $p$ and $q$ are parameters to be determined from the plots of the straight lines in Figs. 2-4; and we provide them in Table 2 below.

For each isotopic chain, the quality of the linear fits is quantified by the *coefficient of determination* $R^2$. Thus, across all elements with $Z<47$, the $R^2$ values cluster near unity, typically exceeding 0.98 and never falling below ≈0.94. It is all the more so, since the $E$ values used in the figures we presented above are available with a precision of about 0.1%, if not even better. No systematic offset is hence observed in the residuals, which indicates the absence of higher-order dependence on $A$ within the examined range.

Note further that, while sophisticated mass models reproduce $\beta$-decay energy values globally, they do not single out or emphasize any strict linear dependence on $A$ at fixed $Z$ as an organizing principle.

**Table 2** The parameters $p$ and $q$ of the straight lines seen in Figs. 2–4 as per $E$ versus $A$ for radioisotopes of the same element.[35] The parity is either *even (e)* or *odd (o)*.

| *Z* | *Stable Nuclei* $(A_{stable})$ | *Radioisotopes* | *Even, Odd* | *p (intercept)* | *q (slope)* | $R^2$ | *-p/q* |
|---|---|---|---|---|---|---|---|
| 0 | | $^{1}$n | *o* | -2.77 | 3.56 | 1 | |
| 1 | | $^{3}$H | *o* | -10.40 | 3.47 | 1 | |
| 2 | $^{3}$He, $^{4}$He | $^{6}$He, $^{8}$He | *e* | -11.27 | 2.67 | 0.96 | 4.22 |
| 3 | $^{6}$Li, $^{7}$Li | $^{8}$Li<br>$^{9}$Li, $^{11}$Li | *e*<br>*o* | -48.01<br>-34.85 | 8.00<br>5.14 | 1<br>0.97 | 6.00<br>6.78 |
| 4 | $^{8}$Be, $^{9}$Be | $^{10}$Be, $^{12}$Be, $^{14}$Be<br>$^{11}$Be | *e*<br>*o* | -25.87<br>-51.79 | 3.00<br>5.75 | 0.91<br>1 | 8.62<br>9.01 |
| 5 | $^{10}$B, $^{11}$B, | $^{12}$B, $^{14}$B, $^{16}$B<br>$^{13}$B, $^{15}$B, $^{17}$B, $^{19}$B | *e*<br>*o* | -36.04<br>-31.49 | 3.88<br>3.20 | 0.91<br>0.92 | 9.29<br>9.84 |
| 6 | $^{12}$C, $^{13}$C | $^{14}$C, $^{16}$C, $^{18}$C, $^{20}$C, $^{22}$C,<br>$^{15}$C, $^{17}$C, $^{19}$C | *e*<br>*o* | -28.90<br>-32.58 | 2.26<br>2.65 | 0.97<br>0.92 | 12.79<br>12.29 |
| 7 | $^{14}$N, $^{15}$N | $^{16}$N, $^{18}$N, $^{20}$N, $^{22}$N, $^{24}$N<br>$^{17}$N, $^{19}$N, $^{21}$N, $^{23}$N | *e*<br>*o* | -34.17<br>-37.98 | 2.62<br>2.64 | 0.97<br>0.98 | 13.04<br>14.39 |
| 8 | $^{18}$O, $^{17}$O | $^{20}$O, $^{22}$O, $^{24}$O<br>$^{19}$O, $^{21}$O, $^{23}$O, $^{25}$O | *e*<br>*o* | -31.96<br>-32.35 | 1.77<br>1.92 | 0.99<br>0.99 | 18.06<br>16.85 |
| 9 | $^{18}$F, $^{19}$F | $^{20}$F, $^{22}$F, $^{24}$F, $^{26}$F, $^{28}$F, $^{30}$F<br>$^{21}$F, $^{23}$F, $^{25}$F, $^{27}$F, $^{29}$F, $^{31}$F | *e*<br>*o* | -35.19<br>-37.40 | 2.05<br>2.01 | 0.99<br>0.99 | 17.16<br>18.61 |
| 10 | $^{22}$Ne, $^{21}$Ne | $^{24}$Ne, $^{26}$Ne, $^{28}$Ne, $^{30}$Ne, $^{32}$Ne, $^{34}$Ne<br>$^{23}$Ne, $^{25}$Ne, $^{27}$Ne, $^{29}$Ne, $^{31}$Ne, $^{33}$Ne | *e*<br>*o* | -40.84<br>-39.02 | 1.85<br>1.88 | 0.99<br>1 | 22.08<br>20.76 |
| 11 | $^{22}$Na, $^{23}$Na | $^{24}$Na, $^{26}$Na, $^{28}$Na, $^{30}$Na, $^{32}$Na, $^{34}$Na, $^{36}$Na<br>$^{25}$Na, $^{27}$Na, $^{29}$Na, , $^{31}$Na, $^{33}$Na, $^{35}$Na,<br>$^{37}$Na | *e*<br>*o* | -37.51<br>-39.76 | 1.79<br>1.77 | 0.98<br>0.99 | 20.96<br>22.46 |
| 12 | $^{25}$Mg, $^{26}$Mg, | $^{28}$Mg, $^{30}$Mg, $^{32}$Mg, $^{34}$Mg, $^{36}$Mg, $^{38}$Mg,<br>$^{40}$Mg<br>$^{27}$Mg, $^{29}$Mg, $^{31}$Mg, $^{33}$Mg, $^{35}$Mg, $^{37}$Mg | *e*<br>*o* | -38.27<br>-38.63 | 1.48<br>1.57 | 0.99<br>0.98 | 25.86<br>24.61 |
| 13 | $^{26}$Al, $^{27}$Al | $^{28}$Al, $^{30}$Al, $^{32}$Al, $^{34}$Al, $^{36}$Al, $^{38}$Al, $^{40}$Al,<br>$^{42}$Al<br>$^{29}$Al, $^{31}$Al, $^{33}$Al, $^{35}$Al, $^{37}$Al, $^{39}$Al, $^{41}$Al,<br>$^{43}$Al | *e*<br>*o* | -38.20<br>-37.99 | 1.55<br>1.46 | 0.97<br>0.98 | 24.65<br>26.02 |
| 14 | $^{29}$Si, $^{30}$Si | $^{32}$Si, $^{34}$Si, $^{36}$Si, $^{38}$Si, $^{40}$Si, $^{42}$Si, $^{44}$Si<br>$^{31}$Si, $^{33}$Si, $^{35}$Si, $^{37}$Si, $^{39}$Si, $^{41}$Si, $^{43}$Si | *e*<br>*o* | -42.45<br>-40.62 | 1.38<br>1.41 | 0.99<br>0.97 | 30.76<br>28.81 |
| 15 | $^{30}$P, $^{31}$P | $^{32}$P, $^{34}$P, $^{36}$P, $^{38}$P, $^{40}$P, $^{42}$P, $^{44}$P, $^{46}$P<br>$^{33}$P, $^{35}$P, $^{37}$P, $^{39}$P, $^{41}$P, $^{43}$P, $^{45}$P, $^{47}$P | *e*<br>*o* | -43.75<br>-46.56 | 1.46<br>1.46 | 0.98<br>0.99 | 29.97<br>31.89 |
| 16 | $^{33}$S, $^{36}$S | $^{38}$S, $^{40}$S, $^{42}$S, $^{44}$S, $^{46}$S, $^{48}$S<br>$^{35}$S, $^{37}$S, $^{39}$S, $^{41}$S, $^{43}$S, $^{45}$S, $^{47}$S, $^{49}$S | *e*<br>*o* | -50.77<br>-45.42 | 1.40<br>1.33 | 0.99<br>0.99 | 36.26<br>34.15 |
| 17 | $^{34}$Cl, $^{37}$Cl | $^{36}$Cl, $^{38}$Cl, $^{40}$Cl, $^{42}$Cl, $^{44}$Cl, $^{46}$Cl, $^{48}$Cl,<br>$^{50}$Cl, $^{52}$Cl<br>$^{39}$Cl, $^{41}$Cl, $^{43}$Cl, $^{45}$Cl, $^{47}$Cl, $^{49}$Cl, $^{51}$Cl | *e*<br>*o* | -47.45<br>-55.54 | 1.37<br>1.50 | 0.99<br>0.99 | 34.64<br>37.03 |
| 18 | $^{37}$Ar, $^{40}$Ar | $^{42}$Ar, $^{44}$Ar, $^{46}$Ar, $^{48}$Ar, $^{50}$Ar, $^{52}$Ar, $^{54}$Ar | *e* | -56.49 | 1.37 | 0.98 | 41.24 |

[35] We used the plotting program OriginPro 2026 (https://www.originlab.com). The *parameters p, q* and the *coefficient of determination* $R^2$ are provided automatically by this program.

| | | | | | | | |
|---|---|---|---|---|---|---|---|
| | | $^{39}$Ar, $^{41}$Ar, $^{43}$Ar, $^{45}$Ar, $^{47}$Ar, $^{49}$Ar, $^{51}$Ar, $^{53}$Ar | *o* | -46.87 | 1.22 | 0.98 | 38.42 |
| 19 | $^{38}$K, $^{41}$K | $^{40}$K, $^{42}$K, $^{44}$K, $^{46}$K, $^{48}$K, $^{50}$K, $^{52}$K, $^{54}$K, $^{56}$K<br>$^{43}$K, $^{45}$K, $^{47}$K, $^{49}$K, $^{51}$K, $^{53}$K, $^{55}$K, $^{57}$K | *e*<br>*o* | -49.86<br>-59.17 | 1.28<br>1.42 | 0.99<br>0.99 | 38.95<br>41.67 |
| 20 | $^{43}$Ca, $^{46}$Ca | $^{48}$Ca, $^{50}$Ca, $^{52}$Ca, $^{54}$Ca, $^{56}$Ca, $^{58}$Ca, $^{60}$Ca<br>$^{45}$Ca, $^{47}$Ca, $^{49}$Ca, $^{51}$Ca, $^{53}$Ca, $^{55}$Ca, $^{57}$Ca, $^{59}$Ca | *e*<br>*o* | -55.68<br>-49.75 | 1.20<br>1.22 | 0.98<br>0.99 | 46.40<br>45.78 |
| 21 | $^{44}$Sc, $^{45}$Sc | $^{46}$Sc, $^{48}$Sc, $^{50}$Sc, $^{52}$Sc, $^{54}$Sc, $^{56}$Sc, $^{58}$Sc, $^{60}$Sc, $^{62}$Sc<br>$^{47}$Sc, $^{49}$Sc, $^{51}$Sc, $^{53}$Sc, $^{55}$Sc, $^{57}$Sc, $^{59}$Sc, $^{61}$Sc | *e*<br>*o* | -47.67<br>-52.48 | 1.09<br>1.14 | 1<br>0.99 | 43.73<br>46.04 |
| 22 | $^{49}$Ti, $^{50}$Ti | $^{52}$Ti, $^{54}$Ti, $^{56}$Ti, $^{58}$Ti, $^{60}$Ti, $^{62}$Ti, $^{64}$Ti<br>$^{51}$Ti, $^{53}$Ti, $^{55}$Ti, $^{57}$Ti, $^{59}$Ti, $^{61}$Ti, $^{63}$Ti | *e*<br>*o* | -53.63<br>-57.05 | 1.08<br>1.17 | 1<br>0.99 | 49.66<br>48.76 |
| 23 | $^{48}$V, $^{51}$V | $^{50}$V, $^{52}$V, $^{54}$V, $^{56}$V, $^{58}$V, $^{60}$V, $^{62}$V, $^{64}$V, $^{66}$V<br>$^{53}$V, $^{55}$V, $^{57}$V, $^{59}$V, $^{61}$V, $^{63}$V, $^{65}$V, $^{67}$V | *e*<br>*o* | -53.16<br>-53.76 | 1.11<br>1.08 | 0.99<br>0.99 | 47.89<br>49.78 |
| 24 | $^{53}$Cr, $^{54}$Cr | $^{56}$Cr, $^{58}$Cr, $^{60}$Cr, $^{62}$Cr, $^{64}$Cr, $^{66}$Cr, $^{68}$Cr, $^{70}$Cr<br>$^{55}$Cr, $^{57}$Cr, $^{59}$Cr, $^{61}$Cr, $^{63}$Cr, $^{65}$Cr, $^{67}$Cr, $^{69}$Cr | *e*<br>*o* | -48.23<br>-52.67 | 0.90<br>1.01 | 0.99<br>0.99 | 53.59<br>52.15 |
| 25 | $^{54}$Mn, $^{55}$Mn | $^{56}$Mn, $^{58}$Mn, $^{60}$Mn, $^{62}$Mn, $^{64}$Mn, $^{66}$Mn, $^{68}$Mn, $^{70}$Mn, $^{72}$Mn<br>$^{57}$Mn, $^{59}$Mn, $^{61}$Mn, $^{63}$Mn, $^{65}$Mn, $^{67}$Mn, $^{69}$Mn, $^{71}$Mn, $^{73}$Mn | *e*<br>*o* | -50.54<br>-49.83 | 0.97<br>0.92 | 0.98<br>0.99 | 52.10<br>54.16 |
| 26 | $^{57}$Fe, $^{58}$Fe | $^{60}$Fe, $^{62}$Fe, $^{64}$Fe, $^{66}$Fe, $^{68}$Fe, $^{70}$Fe, $^{72}$Fe, $^{74}$Fe, $^{76}$Fe<br>$^{59}$Fe, $^{61}$Fe, $^{63}$Fe, $^{65}$Fe, $^{67}$Fe, $^{69}$Fe, $^{71}$Fe, $^{73}$Fe, $^{75}$Fe | *e*<br>*o* | -50.95<br>-49.54 | 0.87<br>0.88 | 0.99<br>0.99 | 58.56<br>56.30 |
| 27 | $^{58}$Co, $^{59}$Co | $^{60}$Co, $^{62}$Co, $^{64}$Co, $^{66}$Co, $^{68}$Co, $^{70}$Co, $^{72}$Co, $^{74}$Co, $^{76}$Co, $^{78}$Co<br>$^{61}$Co, $^{63}$Co, $^{65}$Co, $^{67}$Co, $^{69}$Co, $^{71}$Co, $^{73}$Co, $^{75}$Co, $^{77}$Co | *e*<br>*o* | -52.67<br>-53.56 | 0.93<br>0.91 | 0.99<br>0.99 | 56.53<br>58.86 |
| 28 | $^{61}$Ni, $^{64}$Ni | $^{66}$Ni, $^{68}$Ni, $^{70}$Ni, $^{72}$Ni, $^{74}$Ni, $^{76}$Ni, $^{78}$Ni, $^{80}$Ni<br>$^{63}$Ni, $^{65}$Ni, $^{67}$Ni, $^{69}$Ni, $^{71}$Ni, $^{73}$Ni, $^{75}$Ni, $^{77}$Ni, $^{79}$Ni, $^{81}$Ni | *e*<br>*o* | -54.03<br>-51.40 | 0.83<br>0.83 | 0.99<br>0.99 | 65.10<br>61.93 |
| 29 | $^{64}$Cu, $^{65}$Cu | $^{66}$Cu, $^{68}$Cu, $^{70}$Cu, $^{72}$Cu, $^{74}$Cu, $^{76}$Cu, $^{78}$Cu, $^{80}$Cu, $^{82}$Cu<br>$^{67}$Cu, $^{69}$Cu, $^{71}$Cu, $^{73}$Cu, $^{75}$Cu, $^{77}$Cu, $^{79}$Cu, $^{81}$Cu | *e*<br>*o* | -57.22<br>-59.52 | 0.90<br>0.90 | 1<br>1 | 63.58<br>66.13 |
| 30 | $^{67}$Zn, $^{70}$Zn | $^{72}$Zn, $^{74}$Zn, $^{76}$Zn, $^{78}$Zn, $^{80}$Zn, $^{82}$Zn, $^{84}$Zn<br>$^{69}$Zn, $^{71}$Zn, $^{73}$Zn, $^{75}$Zn, $^{77}$Zn, $^{79}$Zn, $^{81}$Zn, $^{83}$Zn, $^{85}$Zn | *e*<br>*o* | -64.40<br>-55.05 | 0.91<br>0.81 | 0.98<br>1 | 70.77<br>67.96 |
| 31 | $^{68}$Ga, $^{71}$Ga | $^{70}$Ga, $^{72}$Ga, $^{74}$Ga, $^{76}$Ga, $^{78}$Ga, $^{80}$Ga, $^{82}$Ga, $^{84}$Ga, $^{86}$Ga<br>$^{73}$Ga, $^{75}$Ga, $^{77}$Ga, $^{79}$Ga, $^{81}$Ga, $^{83}$Ga, $^{85}$Ga, $^{87}$Ga | *e*<br>*o* | -57.65<br>-66.81 | 0.85<br>0.94 | 1<br>1 | 67.82<br>71.07 |
| 32 | $^{73}$Ge, $^{76}$Ge | $^{78}$Ge, $^{80}$Ge, $^{82}$Ge, $^{84}$Ge, $^{86}$Ge, $^{88}$Ge, $^{90}$Ge<br>$^{75}$Ge, $^{77}$Ge, $^{79}$Ge, $^{81}$Ge, $^{83}$Ge, $^{85}$Ge, $^{87}$Ge, $^{89}$Ge | *e*<br>*o* | -70.79<br>-62.91 | 0.93<br>0.86 | 0.99<br>0.99 | 76.12<br>73.15 |
| 33 | $^{74}$As, $^{75}$As | $^{76}$As, $^{78}$As, $^{80}$As, $^{82}$As, $^{84}$As, $^{86}$As, $^{88}$As, $^{90}$As, $^{92}$As<br>$^{77}$As, $^{79}$As, $^{81}$As, $^{83}$As, $^{85}$As, $^{87}$As, $^{89}$As, $^{91}$As | *e*<br>*o* | -63.72<br>-70.13 | 0.87<br>0.92 | 0.99<br>0.98 | 73.24<br>76.23 |
| 34 | $^{77}$Se, $^{82}$Se | $^{84}$Se, $^{86}$Se, $^{88}$Se, $^{90}$Se, $^{94}$Se<br>$^{79}$Se, $^{81}$Se, $^{83}$Se, $^{85}$Se, $^{87}$Se, $^{89}$Se, $^{91}$Se, $^{93}$Se, $^{95}$Se | *e*<br>*o* | -73.79<br>-63.58 | 0.91<br>0.81 | 0.97<br>0.99 | 81.09<br>78.49 |
| 35 | $^{76}$Br, $^{81}$Br | $^{78}$Br, $^{80}$Br, $^{82}$Br, $^{84}$Br, $^{86}$Br, $^{88}$Br, $^{90}$Br, $^{92}$Br, $^{94}$Br, $^{96}$Br, $^{98}$Br | *e*<br>*o* | -61.04<br>-72.31 | 0.79<br>0.89 | 0.99<br>0.97 | 77.27<br>81.25 |
| | | $^{83}$Br, $^{85}$Br, $^{87}$Br, $^{89}$Br, $^{91}$Br, $^{93}$Br, $^{95}$Br, $^{97}$Br | | | | | |
| 36 | $^{83}$Kr, $^{86}$Kr | $^{88}$Kr, $^{90}$Kr, $^{92}$Kr, $^{94}$Kr, $^{96}$Kr, $^{98}$Kr, $^{100}$Kr<br>$^{85}$Kr, $^{87}$Kr, $^{89}$Kr, $^{91}$Kr, $^{93}$Kr, $^{95}$Kr, $^{97}$Kr, $^{99}$Kr, $^{101}$Kr | *e*<br>*o* | -64.01<br>-64.27 | 076<br>0.78 | 0.98<br>0.99 | 84.22<br>82.40 |
| 37 | $^{84}$Rb, $^{85}$Rb | $^{86}$Rb, $^{88}$Rb, $^{90}$Rb, $^{92}$Rb, $^{94}$Rb, $^{96}$Rb, $^{98}$Rb, $^{100}$Rb, $^{102}$Rb, $^{104}$Rb<br>$^{87}$Rb, $^{89}$Rb, $^{91}$Rb, $^{93}$Rb, $^{95}$Rb, $^{97}$Rb, $^{99}$Rb, $^{101}$Rb, $^{103}$Rb | *e*<br>*o* | -63.89<br>-64.32 | 0.78<br>0.76 | 0.97<br>0.96 | 81.91<br>84.63 |
| 38 | $^{87}$Sr, $^{88}$Sr | $^{90}$Sr, $^{92}$Sr, $^{94}$Sr, $^{96}$Sr, $^{98}$Sr, $^{100}$Sr, $^{102}$Sr, $^{104}$Sr, $^{106}$Sr<br>$^{89}$Sr, $^{91}$Sr, $^{93}$Sr, $^{95}$Sr, $^{97}$Sr, $^{99}$Sr, $^{101}$Sr, $^{103}$Sr, $^{105}$Sr, $^{107}$Sr | *e*<br>*o* | -57.42<br>-59.13 | 0.65<br>0.68 | 0.99<br>1 | 88.34<br>86.96 |
| 39 | $^{88}$Y, $^{89}$Y | $^{90}$Y, $^{92}$Y, $^{94}$Y, $^{96}$Y, $^{98}$Y, $^{100}$Y, $^{102}$Y, $^{104}$Y, $^{106}$Y, $^{108}$Y<br>$^{91}$Y, $^{93}$Y, $^{95}$Y, $^{97}$Y, $^{99}$Y, $^{101}$Y, $^{103}$Y, $^{105}$Y, $^{107}$Y, $^{109}$Y | *e*<br>*o* | -58.97<br>-56.46 | 0.68<br>0.64 | 0.98<br>0.99 | 86.72<br>88.22 |
| 40 | $^{91}$Zr, $^{94}$Zr | $^{96}$Zr, $^{98}$Zr, $^{100}$Zr, $^{102}$Zr, $^{104}$Zr, $^{106}$Zr, $^{108}$Zr, $^{110}$Zr, $^{112}$Zr<br>$^{93}$Zr, $^{95}$Zr, $^{97}$Zr, $^{99}$Zr, $^{101}$Zr, $^{103}$Zr, $^{105}$Zr, $^{107}$Zr, $^{109}$Zr, $^{111}$Zr, $^{113}$Zr | *e*<br>*o* | -57.61<br>-58.61 | 0.61<br>0.63 | 0.99<br>0.99 | 94.44<br>93.03 |
| 41 | $^{92}$Nb, $^{93}$Nb | $^{94}$Nb, $^{96}$Nb, $^{98}$Nb, $^{100}$Nb, $^{102}$Nb, $^{104}$Nb, $^{106}$Nb, $^{108}$Nb, $^{110}$Nb, $^{112}$Nb, $^{114}$Nb, $^{116}$Nb<br>$^{95}$Nb, $^{97}$Nb, $^{99}$Nb, $^{101}$Nb, $^{103}$Nb, $^{105}$Nb, $^{107}$Nb, $^{109}$Nb, $^{111}$Nb, $^{113}$Nb, $^{115}$Nb | *e*<br>*o* | -55.63<br>-58.73 | 0.62<br>0.63 | 1<br>1 | 89.73<br>93.22 |
| 42 | $^{97}$Mo, $^{100}$Mo | $^{102}$Mo, $^{104}$Mo, $^{106}$Mo, $^{108}$Mo, $^{110}$Mo, $^{112}$Mo, $^{114}$Mo, $^{116}$Mo, $^{118}$Mo<br>$^{99}$Mo, $^{101}$Mo, $^{103}$Mo, $^{105}$Mo, $^{107}$Mo, $^{109}$Mo, $^{111}$Mo, $^{113}$Mo, $^{115}$Mo, $^{117}$Mo, $^{119}$Mo | *e*<br>*o* | -63.33<br>-58.16 | 0.63<br>0.60 | 1<br>1 | 100.52<br>96.93 |
| 43 | $^{96}$Tc, $^{97}$Tc | $^{98}$Tc, $^{100}$Tc, $^{102}$c, $^{104}$Tc, $^{106}$Tc, $^{108}$Tc, $^{110}$Tc, $^{112}$Tc, $^{114}$Tc, $^{116}$Tc, $^{118}$Tc, $^{120}$Tc<br>$^{99}$Tc, $^{101}$Tc, $^{103}$c, $^{105}$Tc, $^{107}$Tc, $^{109}$Tc, $^{111}$Tc, $^{113}$Tc, $^{115}$Tc, $^{117}$Tc, $^{119}$Tc, $^{121}$Tc | *e*<br>*o* | -55.22<br>-57.37 | 0.58<br>0.59 | 1<br>1 | 95.21<br>97.24 |
| 44 | $^{101}$Ru, $^{104}$Ru | $^{106}$Ru, $^{108}$Ru, $^{110}$Ru, $^{112}$Ru, $^{114}$Ru, $^{116}$Ru, $^{118}$Ru, $^{120}$Ru, $^{122}$Ru, $^{124}$Ru<br>$^{103}$Ru, $^{105}$Ru, $^{107}$Ru, $^{109}$Ru, $^{111}$Ru, $^{113}$Ru, $^{115}$Ru, $^{117}$Ru, $^{119}$Ru, $^{121}$Ru, $^{123}$Ru, $^{125}$Ru | *e*<br>*o* | -60.71<br>-55.87 | 0.58<br>0.55 | 0.99<br>0.99 | 104.67<br>101.58 |
| 45 | $^{102}$Rh, $^{103}$Rh | $^{104}$Rh, $^{106}$Rh, $^{108}$Rh, $^{110}$Rh, $^{112}$Rh, $^{114}$Rh, $^{116}$Rh, $^{118}$Rh, $^{120}$Rh, $^{122}$Rh, $^{124}$Rh, $^{126}$Rh, $^{128}$Rh<br>$^{105}$Rh, $^{107}$Rh, $^{109}$Rh, $^{111}$Rh, $^{113}$Rh, $^{115}$Rh, $^{117}$Rh, $^{119}$Rh, $^{121}$Rh, $^{123}$Rh, $^{125}$Rh, $^{127}$Rh | *e*<br>*o* | -60.19<br>-59.05 | 0.60<br>0.57 | 0.99<br>1 | 100.32<br>103.60 |
| 46 | $^{105}$Pd, $^{110}$Pd | $^{112}$Pd, $^{114}$Pd, $^{116}$Pd, $^{118}$Pd, $^{120}$Pd, $^{122}$Pd, $^{124}$Pd, $^{126}$Pd, $^{128}$Pd, $^{130}$Pd<br>$^{107}$Pd, $^{109}$Pd, $^{111}$Pd, $^{113}$Pd, $^{115}$Pd, $^{117}$Pd, $^{119}$Pd, $^{121}$Pd, $^{123}$Pd, $^{125}$Pd, $^{127}$Pd, $^{129}$Pd, $^{131}$Pd | *e*<br>*o* | -71.08<br>-62.18 | 0.64<br>0.58 | 0.99<br>0.99 | 111.06<br>107.21 |

It would certainly be interesting if we could formalize the relationship between the parameters $p$ and $q$ the way presented in Table 2 above. For one thing, as we will show forthrightly, they depend on each other.

To achieve this, we rewrite Eq. (4) for a given $Z$ as

$$E= p+qA+qA_{stable}-qA_{stable}\,, \quad (2)$$

or as

$$E=p+ q(A-A_{stable})+qA_{stable}\,. \quad (3)$$

Let us pose:

$$\Delta A = A - A_{stable} \quad , \tag{4}$$

as a positive quantity.

Let us furthermore transcribe

$$Q = p + qA_{stable}. \tag{5}$$

This leads to

$$E = q\Delta A + Q \quad . \tag{6}$$

It is evident that, for $\Delta A=0$, meaning for the stable nucleus, $E$ will be 0. Therefore,

$$Q = 0 \ , \tag{7}$$

or, the same,

$$-\frac{p}{q} = A_{stable} \quad . \tag{8}$$

We prepared Fig. 5 below to numerically cross-check this *correspondence,* where $-p/q$ values from Table 2 are plotted against the corresponding $A_{stable}$, which we expect to trace the bisector of the coordinate system; and it remarkably does!

Thus,

$$E = q\Delta A \quad , \tag{9}$$

for a given $Z$.[36]

This is an unexpectedly elegant relationship in full conformity with the Figs. 1, 2, 3 and 4 presented above.

We are thereby left with just one parameter—*i.e., q*—which is easy to determine; and it is already furnished in the last column of Table 2 above.

---

[36] The Universal Matter Architecture (UMA) holds valid for the entire body of quantum mechanics; it allows us to perceive how different parameters—*i.e., mass, period of time, size and energy*—associated with a given entity are interrelated with each other, and this can be shown to be true without the need to solve the cumbersome quantum mechanical description of the said entity. One other thing is that we can check every aspect of UMA on the Hydrogen (H) atom as the simplest baseline. While we referred to UMA in several places so far just to recall the fact that it was basically the tool which motivated us to draw the linear dependence of $E$ on $A$ (cf. Figs 1-4), the corresponding mathematical details lie far beyond the content of this article. Still, it will be enriching to check Eq. (9), on the basis of H atom. Thus, we consider an outmost and practically stable energy level of the H atom determined by the quantum number $n$. We can now express the stability of this level with respect to an even upper energy level close to the previous one as determined by the quantum number $n+\Delta n$ via writing the Bohr decay-energy $E_{k\to N}=(2\pi^2e^4m/h^2)[1/n^2-1/(n+\Delta n)^2]$, where $e$ is the electron charge, $m$ the electron mass, and $h$ the Planck Constant. If $n$ *is* sufficiently big, this can be written as $E_{k\to N}=2C[\Delta n/n^3]$, where we have $C=2\pi^2e^4m/h^2$. Via defining a new constant, $C'=2C/n^3$, we thus recover the cast of Eq. (9) finally written as $E_{k\to N}=C'\Delta n$. The analogy we depicted is solely intended to illustrate how linear quantization can naturally emerge in bound quantum systems without implying a full physical equivalence between atomic and nuclear dynamics. All the same, the present nuclear results provide an empirical arena by which the broader applicability of the UMA can be meaningfully addressed.

It so happens that, in the case of radioisotopes of the same element, knowing just one $\beta^-$-decay energy $E$ for a given $A$ of elongation $\Delta A$ to the stable nucleus $A_{stable}$, one can write

$$q = \frac{E}{\Delta A} \quad (10)$$

To corroborate this result, we plot in Figs. 6a, 6b and 6c the $E_{measured}$ versus $E_{calculated} = q\Delta A$ quantities for the elements Cobalt, Gallium and Molybdenum that pertain, respectively, to $Z$=27, $Z$=31 and $Z$=42.

The drawings are based on data from Tables 3a, 3b and 3c in succession; and the fact that the final outcome yields a neat alignment with the bisector every time is a witness to the explanatory and predictive power of the UMA scaffolding.

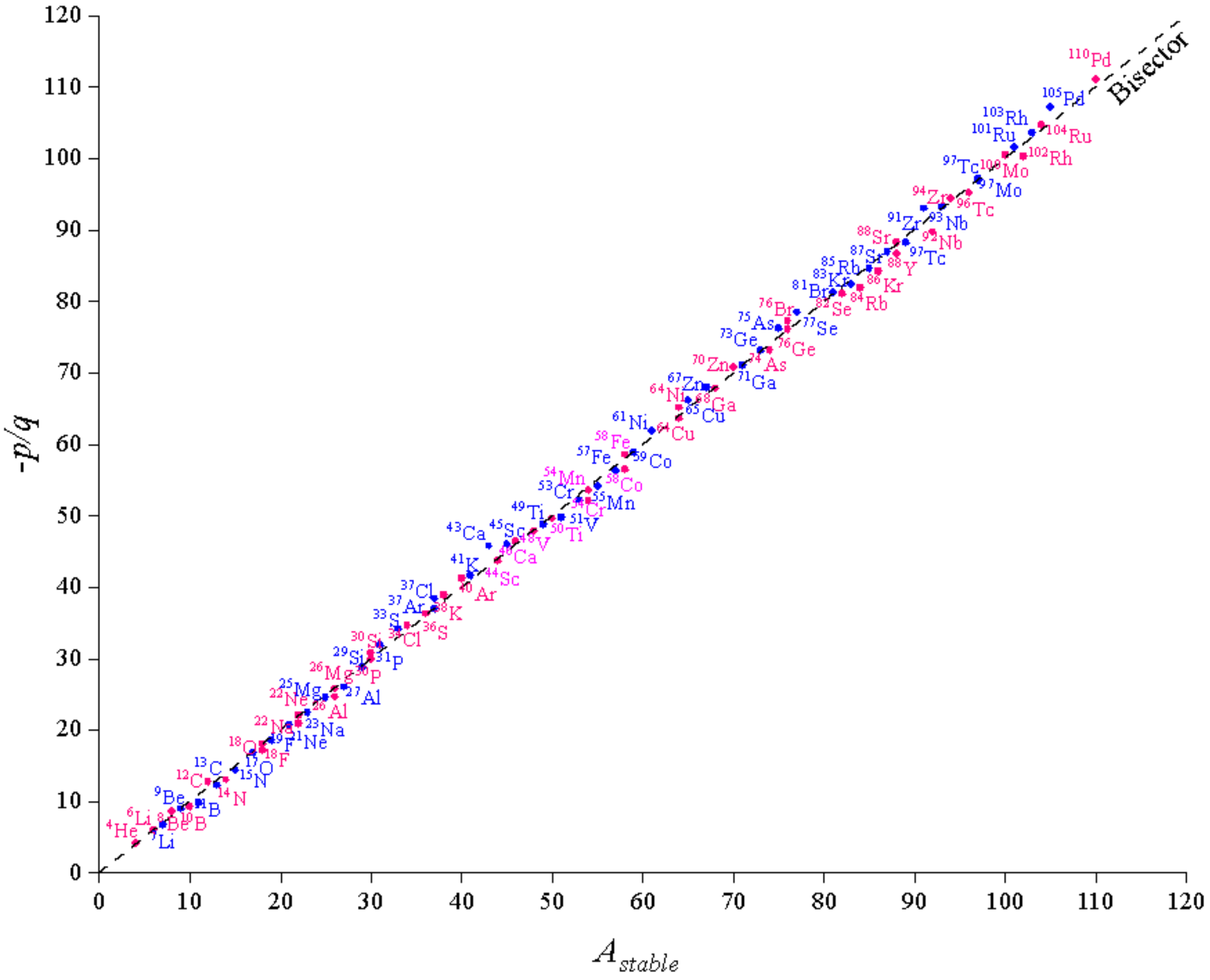

**Figure 5** The $-p/q$ relationship from $E=p+qA$ (cf. Figs. 2, 3 and 4) versus $A_{stable}$ in accordance with Eq. (10). When it is possible to find more than one stable nucleus for a given element, the stable nucleus nearest to the $\beta^-$-decaying nuclei set is considered. All nuclei are then expected to be situated on the bisector of the graph which, very satisfactorily, is indeed the case.

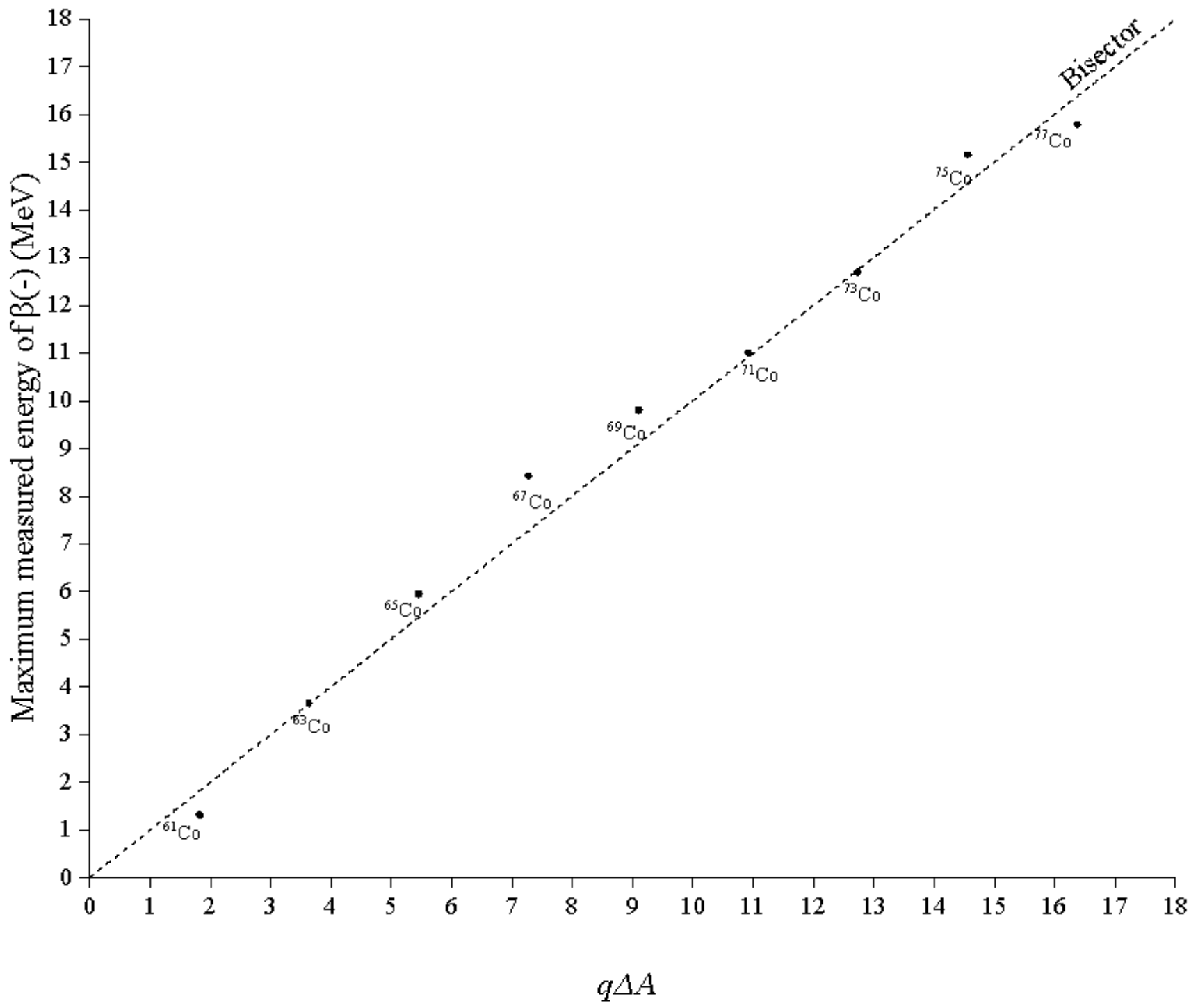

**Figure 6a** $E_{measured}$ versus $E_{calculated}=q\Delta A$ for Cobalt.

**Table 3a** $E_{measured}$ versus $E_{calculated}=q\Delta A$ for Cobalt, where $A_{stable}$=59 and $q$=0.91.

| *Radioisotopes* | *ΔA* | *qΔA* | *E (MeV)* |
|---|---|---|---|
| $^{61}Co$ | 2 | 1.82 | 1.32 |
| $^{63}Co$ | 4 | 3.64 | 3.66 |
| $^{65}Co$ | 6 | 5.46 | 5.94 |
| $^{67}Co$ | 8 | 7.28 | 8.42 |
| $^{69}Co$ | 10 | 9.10 | 9.81 |
| $^{71}Co$ | 12 | 10.92 | 11.0 |
| $^{73}Co$ | 14 | 12.74 | 12.69 |
| $^{75}Co$ | 16 | 14.56 | 15.15 |
| $^{77}Co$ | 18 | 16.38 | 15.79 |

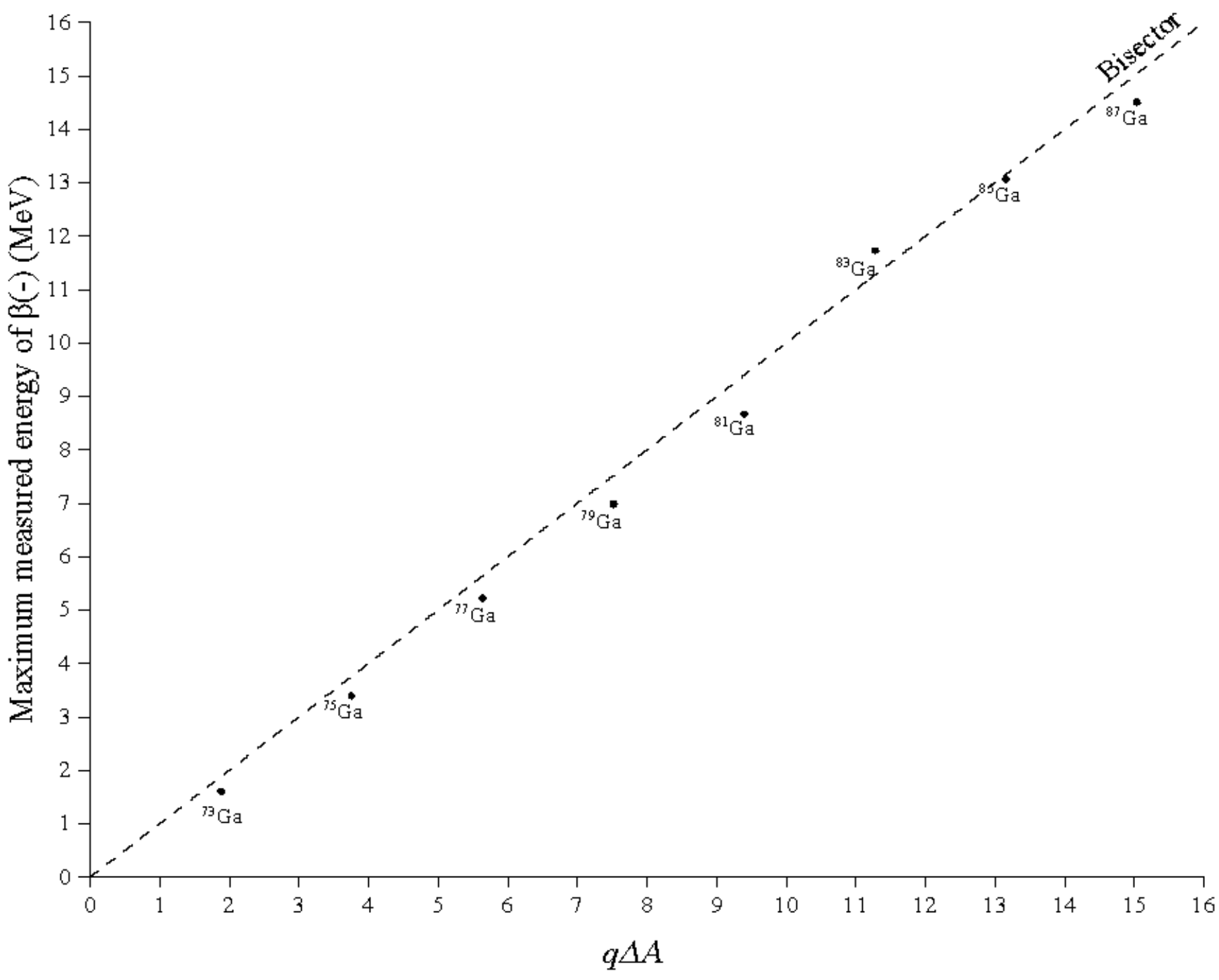

**Figure 6b** $E_{measured}$ versus $E_{calculated}=q\Delta A$ for Gallium.

**Table 3b** $E_{measured}$ versus $E_{calculated}=q\Delta A$ for Gallium, where $A_{stable}$=71 and $q$=0.94.

| *Radioisotopes* | $\Delta A$ | $q\Delta A$ | *E (MeV)* |
|---|---|---|---|
| $^{73}Ga$ | 2 | 1.88 | 1.6 |
| $^{75}Ga$ | 4 | 3.76 | 3.39 |
| $^{77}Ga$ | 6 | 5.64 | 5.22 |
| $^{79}Ga$ | 8 | 7.52 | 6.98 |
| $^{81}Ga$ | 10 | 9.40 | 8.66 |
| $^{83}Ga$ | 12 | 11.28 | 11.72 |
| $^{85}Ga$ | 14 | 13.16 | 13.06 |
| $^{87}Ga$ | 16 | 15.04 | 14.50 |

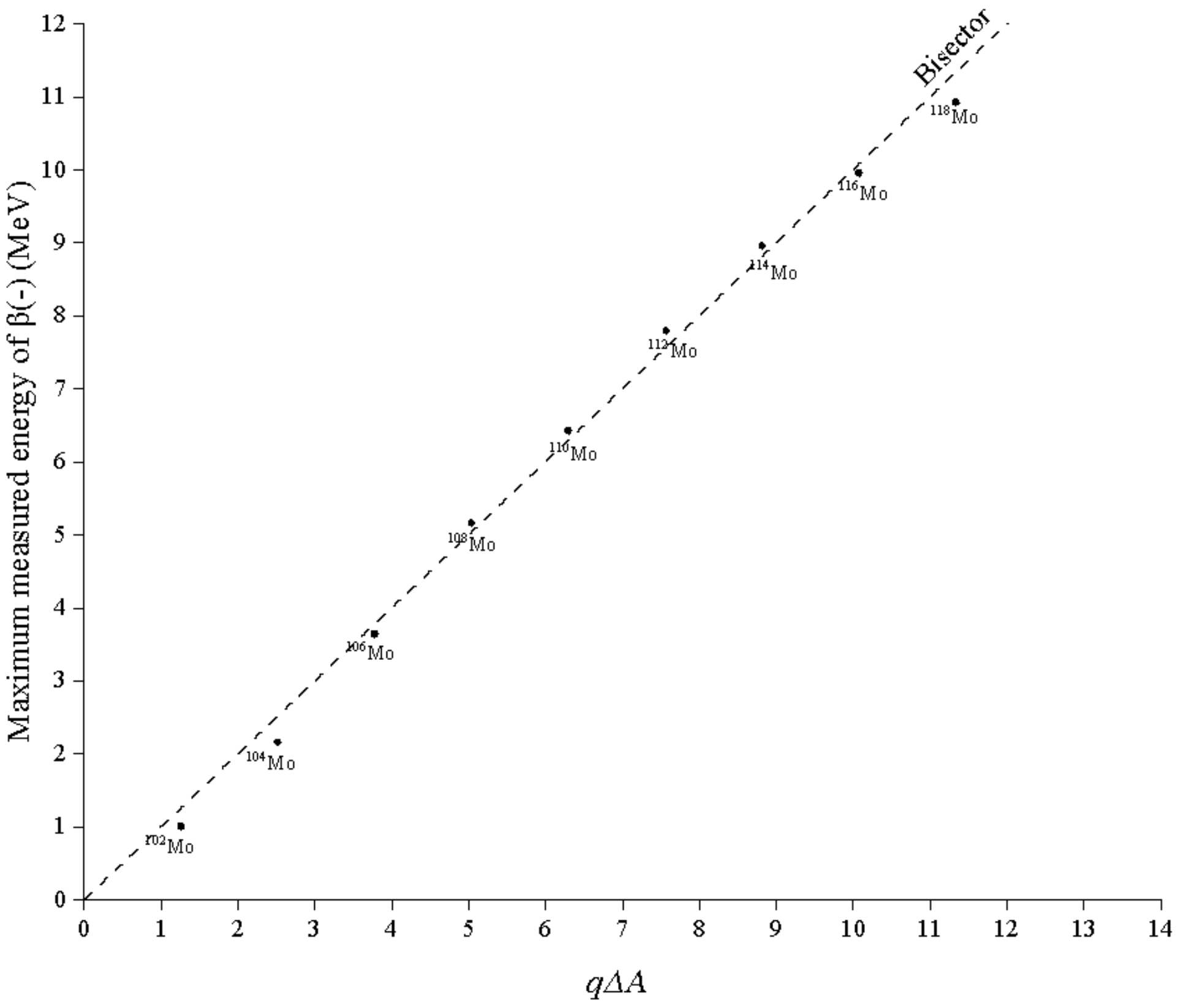

**Figure 6c** $E_{measured}$ versus $E_{calculated}=q\Delta A$ for Molybdenum.

**Table 3c** $E_{measured}$ versus $E_{calculated}=q\Delta A$ for Molybdenum, where $A_{stable}=100$ and $q=0.60$.

| *Radioisotope* | $\Delta A$ | $E_{calculated}=q\Delta A$ *(MeV)* | $E_{measured}$ *(MeV)* |
|---|---|---|---|
| $^{102}Mo$ | 2 | 1.26 | 1.01 |
| $^{104}Mo$ | 4 | 2.52 | 2.16 |
| $^{106}Mo$ | 6 | 3.78 | 3.64 |
| $^{108}Mo$ | 8 | 5.04 | 5.16 |
| $^{110}Mo$ | 10 | 6.30 | 6.43 |
| $^{112}Mo$ | 12 | 7.57 | 7.79 |
| $^{114}Mo$ | 14 | 8.82 | 8.96 |
| $^{116}Mo$ | 16 | 10.08 | 9.96 |
| $^{118}Mo$ | 18 | 11.34 | 10.92 |

As the above Figs. 6a, 6b and 6c witness, Eq. (9) confirms the generic quantization of $\beta^-$-decay maximum energy $E$ of the nucleus $A$ within the given element $Z$. This approach thus uniquely frames how *$\beta^-$-decay nuclear instability ought to be quantized.*

One key point to be emphasized is that, had we endeavored to express $E$ as a function of $A$, this function, owing to the UMA, *and hence in the light of the preceding discussion,* would be linear. It is because, as discussed above, the behavior in question appears to be a quantum mechanical necessity.

More specifically, the *linearity* under consideration quantifies the *increase in the instability* of $\beta^-$-decay as one jumps from $A$ to $A$+1 within the given element. In different terms, what we have here is *a unique quantization of nuclear instability.*

Besides this, note that, in the UMA framework, the $\beta^-$-decay lifetime is governed by the effective energy release $E$ through the generic scaling [29, 30]

$$T \propto \sqrt{\frac{1}{E}} \quad . \tag{11}$$

We would like to accentuate that this scaling, in relation to decay half-lives, serves here only as a phenomenological guide for interpreting the observed linear energy $E$ with respect $A$ for nuclei configured practically alike.

## 4. CONCLUSION

In this work, we have demonstrated that the maximum $\beta^-$-decay energy $E$ exhibits a remarkably simple and robust linear dependence on the mass number $A$ with respect to any particular isotopic chain of fixed proton number $Z$ in the range $Z$<47 encompassed by this study, provided that even-$A$ and odd-$A$ nuclei are treated separately.

Thus, across all the elements examined herein, the curated nuclear data are matched with excellent accuracy by two straight-line trends that are characterized by the slope and intercept parameters in the way we have systematically tabulated so far.

The quality of these linear relations, which are reflected in the coefficients of determination being typically very close to unity, indicates that the observed behavior is not incidental, but represents a genuine and previously unrecognized regularity in nuclear systematics.

While approximate linear trends may be expected from smooth mass formulas, the present work demonstrates a remarkably precise and systematic organization of the data into linear relations across a wide range of isotopic chains, as predicted in effect by UMA *(cf. the footnote below Eq. (9)).*

The simplicity of our proposed description stands in contrast to the complexity of existing models, and provides a compact framework for organizing $\beta^-$-decay energetics along isotopic chains.

Beyond its conceptual interest to the scientific community, the present approach lends itself as a practical tool for analyzing decay-energy trends and for making preliminary estimates of $\beta^-$-decay properties in regions where data are sparse or incomplete. It may also serve as a useful benchmark for testing and constraining more detailed nuclear-structure models.

The broader theoretical considerations that originally motivated the search for the disclosed regularity lie outside the scope of the present contribution. The linear dependence reported here can be assessed independently on the basis of the experimental data alone.

Be that as it may, in a subsequent paper, we aim to present a similar study for $\beta^-$-decaying nuclei of the radioisotopes of the same element in the range $Z$>46.

## DATA AVAILABILITY

All datasets referenced are publicly available from the cited collaborations.